\documentclass[preprint,nofootinbib,aps,showpacs,11pt]{revtex4-1}
\usepackage{booktabs}
\usepackage{array}
\usepackage{slashed}
\usepackage{braket}
\usepackage{multirow}
\usepackage{mathrsfs}
\usepackage{amsmath}
\usepackage{amssymb}
\usepackage{graphicx}
\usepackage{float}
\usepackage[left]{lineno}
\usepackage{color}
\usepackage{wrapfig}
\usepackage{relsize}
\usepackage{epstopdf}
\usepackage[colorlinks=true,linkcolor=red,citecolor=blue]{hyperref}

\allowdisplaybreaks[1]

\def\orcid#1{\kern .08em\href{https://orcid.org/#1}{\includegraphics[keepaspectratio,width=0.76em]{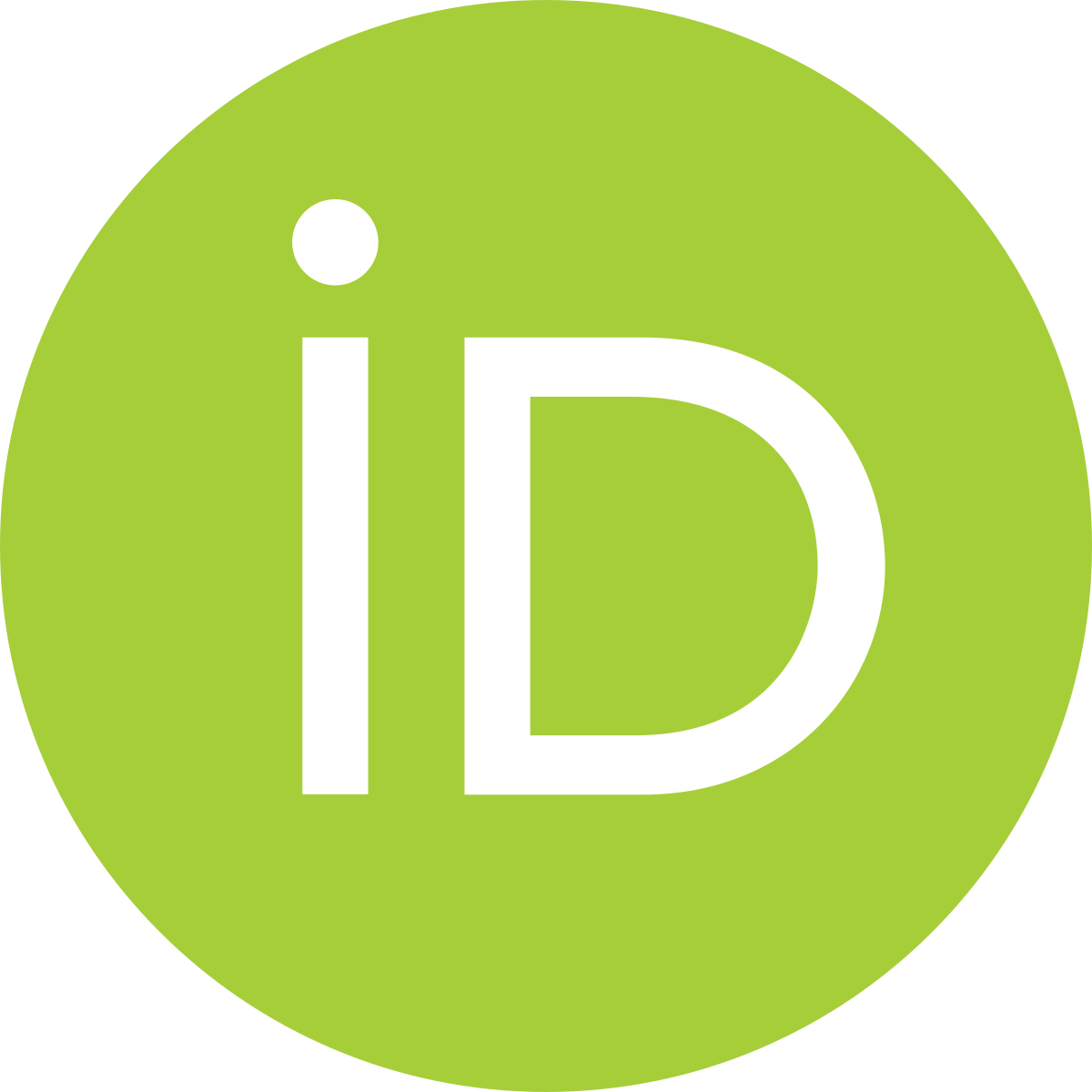}}}
\newcommand{\beq}{\begin{eqnarray}}
\newcommand{\eeq}{\end{eqnarray}}
\newcommand{\non}{\nonumber\\ }

\newcommand{\psl}{ P \hspace{-2.8truemm}/ }

\newcommand{\epsl}{\epsilon \hspace{-1.6truemm}/\,  }

\def\lsim{ {\ \lower-1.2pt\vbox{\hbox{\rlap{$<$}\lower6pt\vbox{\hbox{$\sim$}
  		}}}\ } }
\def\gsim{ {\ \lower-1.2pt\vbox{\hbox{\rlap{$>$}\lower6pt\vbox{\hbox{$\sim$}
  		}}}\ } }

\begin{document}
\title{\boldmath Studying the $B^{0} \to J/\psi h_{1}$ decays with $h_{1}(1170)-h_{1}(1415)$ mixing in the perturbative QCD approach}

\author{
	Qin Chang$^{1,2}$ \footnote{changqin@htu.edu.cn},
	De-Hua Yao$^1$ \footnote{yaodehua@stu.htu.edu.cn},
	Xin~Liu$^3$ \footnote{liuxin@jsnu.edu.cn}
}

\affiliation{
	1.~Centre for Theoretical Physics, Henan Normal University, Xinxiang 453007, China\\
	2.~Center for High Energy Physics, Henan Academy of Sciences, Zhengzhou 455004, China\\
	3.~Department of Physics, Jiangsu Normal University, Xuzhou 221116, China
}

\begin{abstract}
In this paper, we study
the $B^{0} \to J/\psi h_{1}$ decays for the first time by using
perturbative QCD approach up to the presently known next-to-leading order accuracy. 	
The vertex corrections present significant contribution to the amplitude.
In the calculation, the mixing between two light axial-vector mesons $h_{1}(1170)$
and $h_{1}(1415)$ are also studied in detail. The observables including the branching 
ratios, polarization fractions and {\it CP} asymmetries are predicted and discussed explicitly.
It is found that the $B^{0} \to J/\psi h_{1}$ decays have relatively large branching
fractions, which are generally at the order of ${\cal O}(10^{-6}\sim10^{-3})$, and thus are
possible to be observed by the LHCb and Belle-II experiments in the near future. Moreover,
they are very sensitive to the mixing angle $\theta$ and can be used to test the values of $\theta$.
In addition, some ratios between the branching fractions of $B^{0} \to J/\psi h_{1}$ decays
can provide much stronger constraints on $\theta$ due to their relatively small theoretical errors.
The $B^{0} \to J/\psi h_{1}$ decays are generally dominated by the longitudinal polarization contributions, specifically,
$f_{L}(B^{0} \to J/\psi h_{1})>80\%$, except for the case that $\theta\sim  35^\circ$
and $-55^\circ$. Unfortunately, the direct {\it CP} asymmetries of $B^{0} \to J/\psi h_{1}$ decays
are too small to be observed soon even if the effect of $\theta$ is considered. The future precise
measurements on $B^{0} \to J/\psi h_{1}$ decays are expected for testing these theoretical
findings and exploring the interesting nature of $h_{1}(1170)$ and $h_{1}(1415)$.
\end{abstract}
\maketitle

\section{Introduction}
\label{sec01}
It has been known that the decays of $B$ meson
are highly important for exploring the {\it CP} violation, which is expected to be
helpful in the search of
new physics beyond the Standard Model (SM) potentially.
Particularly, the exclusive decays of $B$ meson into a charmonium plus light hadrons have
absorbed a lot of attention in the past decades because
they play a special role in the studies of $B^0-\bar{B^0}$ mixing phase and associated {\it CP}
violation~\cite{Belle:2001zzw,Belle:2001rjp},
while our understanding on their decay mechanism is still far from satisfactory,
though lots of efforts have been made to investigate these decay modes.

In the quark model, according to different spin degeneracy, the $p$-wave axial-vector
meson contains two types of different spectroscopic notation, namely,
$n^{2S+1}L_{J}=1^{3}P_{1}$ and $1^{1}P_{1}$ corresponding to $J^{PC} = 1^{++}$ and $1^{+-}$,
respectively. Intuitively, it is obvious that these two nonets have distinguishable
quantum number {\it C} for the corresponding neutral mesons, i.e.,
{\it C} $=+$ and {\it C} $=-$~\cite{Burakovsky:1997dd,Cheng:2011pb,Chen:2015iqa}.
Experimentally, the $1^{+-}$ multiplets comprise
$b_{1}(1235) (b_1), h_1 [h_{1}(1170),h_{1}(1415)]$ and $K_{1B}$, while the $1^{++}$ multiplets comprise
$a_{1}(1260) (a_1), f_1 [f_{1}(1285),f_{1}(1420)]$ and $K_{1A}$~\cite{ParticleDataGroup:2022pth,HFLAV:2022esi}.
Some efforts
have been made to study these light unflavored axial-vectors~\cite{Liang:2019vhf,
Liu:2010epa,Liu:2010da,Du:2021zdg,Du:2022nno}.
The considered light axial-vector $h_{1}$ states, namely, $h_{1}(1170)$ and $h_{1}(1415)$,
are the important subjects of numerous experimental measurements over the past
decade~\cite{BESIII:2015vfb,BESIII:2018ede,BESIII:2022zel}.
Nevertheless, the Particle Data Group (PDG)~\cite{ParticleDataGroup:2024cfk} continues to report ``No data'' on the
associated decay modes for the $h_{1}$ states.
Theoretically, similar to the $\eta-\eta^{\prime}$ mixing in the pseudoscalar sector,
the mixing scheme of the two physical $h_{1}$ mesons in the singlet-octet (SO) and quark flavor (QF)
basis, respectively, can be written as~\cite{Cheng:2011pb,Cheng:2013cwa}
\beq
\left(
\begin{array}{c} h_{1}(1170)\\ h_{1}(1415) \\ \end{array} \right) =
\left( \begin{array}{cc}
 \,\,\cos{\theta} & \,\,\sin{\theta} \\
 -\sin{\theta} & \,\,\cos{\theta} \end{array} \right)
\left( \begin{array}{c}  \tilde{h}_{1}\\ \tilde{h}_{8} \\ \end{array} \right) =
\left( \begin{array}{cc}
 \,\,\cos{\alpha} & \,\,\sin{\alpha} \\
 -\sin{\alpha} & \,\,\cos{\alpha} \end{array} \right)
\left( \begin{array}{c}  h_{n}\\ h_{s} \\ \end{array} \right)\;,
\label{eq:mix-SO-QF}
\eeq
where the SO states $\tilde{h}_{1}\equiv(u\bar{u}+d\bar{d}+s\bar{s})/\sqrt{3}$,
$\tilde{h}_{8}\equiv(u\bar{u}+d\bar{d}-2s\bar{s})/\sqrt{6}$, and the QF states
$h_{n}\equiv(u\bar{u}+d\bar{d})/\sqrt{2}$, $h_{s}\equiv s\bar{s}$, and their mixing angles
$\theta$ and $\alpha$ obey the following relation~\cite{ParticleDataGroup:2022pth,Du:2021zdg},
\beq
\theta=\alpha-{\rm arctan}\sqrt{2} \,.
\label{eq:angle-relation}
\eeq
Therefore, on the one hand, a profound understanding of the mixing angle $\theta$($\alpha$)
could provide further insight into the hadronic structure of $h_{1}$.
On the other hand, with the help of Gell-Mann$-$Okubo (GMO) relations, the mixing angles
between $h_{1}$ mesons, and between $f_{1}$ 
mesons
have the potential to constrain the mixing angle $\theta_{K}$ between $^3P_1$ state $K_{1A}$
and $^1P_1$ one $K_{1B}$~\cite{Cheng:2011pb,Du:2021zdg}, and vice versa.
It means that an effective constraint on the mixing angle of $h_{1}$ mesons can indirectly
control the range of $\theta_{K}$~\cite{Burakovsky:1997dd,Suzuki:1993yc,Blundell:1995au,Burakovsky:1997ci,
Dag:2012zz,Cheng:2013cwa,Divotgey:2013jba,Gao:2019jme}, which is helpful for analyzing the nature of
$K_{1}(1270)$ and $K_{1}(1400)$. Given that the essential
parameters of
$\tilde{h}_{1}$ and $\tilde{h}_{8}$ have been provided
in the SO mixing scheme from the hadron physics side~\cite{Yang:2007zt}, one can perform
a systematic investigation of relevant $B$-meson weak
decays involving these mentioned states.

\begin{figure}[!!t]
\centering
\begin{tabular}{l}
\includegraphics[width=0.90\textwidth]{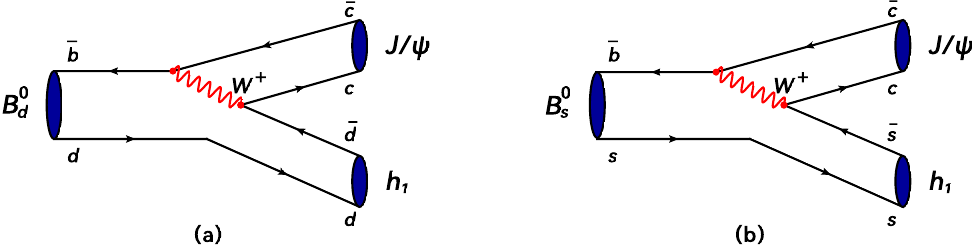}
\end{tabular}
\caption{(Color online) Leading quark-level Feynman diagrams for $B^0 \to J/\psi h_{1}$
}
\label{fig:fig1}
\end{figure}

In this article, we will systematically study
the $B^{0} \to J/\psi h_{1}$ decays~($B^{0}$ includes $B_{d}^{0}$ and $B_{s}^{0}$),
and further search for new observables for determining
the mixing angle $\theta$ in $h_1$ states.
The corresponding transition processes at the quark level
are illustrated in Fig.~\ref{fig:fig1}.
Several groups have studied the $B$-meson decays into a charmonium
state~\cite{Chen:2005ht,Li:2007xf,Li:2012sw,Liu:2014doa,Yao:2022zom,Liu:2019ymi,
Cheng:2000kt,Sun:2013dla,Song:2002gw,Meng:2005er,Li:2006vq,Beneke:2008pi,Colangelo:2010wg,
Liu:2013nea,Wang:2015uea,Zhang:2017cbi,Rui:2019yxx,Li:2020app},
and found that, in order to obtain reliable theoretical predictions for
these color-suppressed transitions that are compatible with the current data,
the next-to-leading order (NLO) QCD corrections, especially the vertex
contributions, and the associated NLO Wilson coefficients, have to be
taken into account in the related calculations.
Besides, it is worth mentioning that the
Sudakov factor for charmonia has been derived recently~\cite{Liu:2018kuo,Liu:2020upy,Liu:2023kxr}
and will also be taken into account in this work.
Combined with the above new ingredients, we will provide
theoretical predictions for the first time on several observables
of $B^0 \to J/\psi h_{1}$, e.g., {\it CP}-averaged branching ratios,
polarization fractions, relative phases, {\it CP}-violating asymmetries, and so on, by
using the PQCD approach~\cite{Keum:2000ph,Keum:2000wi,Lu:2000em,Lu:2000hj,Ali:2007ff,Chai:2022ptk}
up to NLO accuracy. Comparing with the experimental data, the branching ratios
and several interesting ratios could help us to effectively constrain the range of
mixing angle $\theta$, and can provide more useful information for identifying
the inner structure of the $h_{1}$ states.

This paper is organized as follows. In Sect.~\ref{sec02}, we
overview briefly the mixing angle of axial-vector $h_{1}$ mesons
indirectly through the Gell-Mann$-$Okubo mass relations, and then
present the analytic expressions for the $B^{0} \to J/\psi h_{1}$
decay amplitudes in the PQCD approach. In Sect.~\ref{sec03}, the
values of requisite input parameters are collected,
the theoretical results are given and the phenomenological
discussions are made in detail. Finally, we give our summary in Sect.~\ref{sec04}.

\section{Formalism and Perturbative QCD calculations}
\label{sec02}
\subsection{ The mixing angle}
\label{sec0201}
It is known that the physical eigenstates  $K_{1}(1270)$ and $K_{1}(1400)$ are treated as the mixtures
of $K_{1A}$ and $K_{1B}$, which can be expressed as~\cite{Suzuki:1993yc}
\beq
\left(
\begin{array}{c} K_{1}(1270)\\ K_{1}(1400) \\ \end{array} \right) =
\left( \begin{array}{cc}
    \sin{\theta}_{K} & \,\,\cos{\theta}_{K} \\
    \cos{\theta}_{K} & -\sin{\theta}_{K} \end{array} \right)
\left( \begin{array}{c}  {K}_{1A}\\ {K}_{1B} \\
\end{array} \right) \;,
\label{eq:mix-K1}
\eeq
where $\theta_{K}$ is the mixing angle.

\begin{wraptable}{r}{8cm}
\vspace{-1.0cm}
\renewcommand{\arraystretch}{1.20}
\caption{ Theoretical predictions for $\theta_{K}$ }
\label{tab:theory angle}
\begin{center}\vspace{-0.5cm}
\begin{tabular}[t]{r|l}
\hline \hline
\makebox[0.21\textwidth][c]{Models}
&\makebox[0.25\textwidth][c]{  Results of $\theta_{K}$ }
\\ \hline \hline
\multirow{2}{*}{SO basis~\cite{Cheng:2013cwa,Gao:2019jme} }
& $\theta_{K}=33^{\circ}$ \\
& $\theta_{K}=(42.6\pm2.2)^{\circ}$ \\ \hline
\multirow{2}{*}{QF basis~\cite{Suzuki:1993yc,Divotgey:2013jba} }
& $\theta_{K}\approx 33^{\circ}$ \\
& $\theta_{K}=(33.6\pm4.3)^{\circ}$ \\  \hline
\multirow{2}{*}{NRQM~\cite{Burakovsky:1997dd,Burakovsky:1997ci} }
& $35^{\circ}\leq\theta_{K}\leq55^{\circ}$ \\
& $\theta_{K}=(37.3 \pm 3.2)^{\circ}$ \\  \hline
QCD Sum Rules~\cite{Dag:2012zz}
& $\theta_{K}=(39 \pm 4)^{\circ}$ \\ \hline
Relativized QM~\cite{Blundell:1995au}
& $\theta_{K}\approx45^{\circ}$ \\  \hline
{ \large {\bf Average}} & $\theta_{K} \approx 39^{\circ}$ \\
\hline \hline
\end{tabular}
\end{center}
\vspace{-0.5cm}
\end{wraptable}

The value of mixing angle $\theta_{K}$ has been studied in the previous works.
Using the early experimental information of $\tau$ decays, the authors of
Ref.~\cite{Suzuki:1993yc} obtain two solutions $\theta_{K} \approx 33^{\circ}$ or $57^{\circ}$,
and present that the observed $K_1(1400)$ production dominance in the $\tau$ decay favors
$\theta_{K} \approx 33^{\circ}$, while $\theta_{K}=(42.6 \pm 2.2)^{\circ}$ is obtained
in Ref.~\cite{Gao:2019jme} by using BESIII measured $M_{h_1(1415) }= (1423.2 \pm 7.6)\,{\rm MeV}$,
which is larger than the last BESIII measurement  $M_{h_1(1415) }= (1384 \pm 6)\,{\rm MeV}$~\cite{BESIII:2022zel}.
The phenomenological analysis of the $\tau \to K_{1} \nu_{\tau}$ decay suggested that
$\theta_{K} \approx 45^{\circ}$~\cite{Blundell:1995au} in the relativized quark model.
In Ref.~\cite{Cheng:2013cwa}, the author again reinforces the statement that a relatively
small $\theta_{K}=33^{\circ}$ is much more favored by the lattice and phenomenological analyses.
In the nonrelativistic quark model, the range $35^{\circ}\leq \theta_{K} \leq 55^{\circ}$~\cite{Burakovsky:1997dd}
is obtained, and a refined result $\theta_{K}=(37.3\pm3.2)^{\circ}$~\cite{Burakovsky:1997ci}
is given by using the masses of $b_{1}(1235)$ and $a_{1}(1260)$ mesons.
A similar result $\theta_{K}=(39 \pm 4)^{\circ}$~\cite{Dag:2012zz} is obtained by calculating
a two-point correlation function related to $\theta_{K}$ within QCD sum rules. 
The results mentioned above are collected in Table~\ref{tab:theory angle}, and their average
value is $\theta_{K}=39^{\circ}$. Then, we would like to clarify the way to obtain the value
of mixing angle $\theta$ by using $\theta_K$.

Applying the Gell-Mann$-$Okubo relations (specific pedagogical conclusions can be found in
Appendix~\ref{sec:app1}), the mass squared of the octet state can be written as
\beq
m_{\tilde{h}_{8}}^{2}=\frac{4m_{K_{1B}}^{2}-m_{b_{1}}^{2}}{3} \;,
\eeq
where $K_{1B}$ and $b_{1}$ are light meson states belonging to the $^{1}P_{1}$ nonet.
The mixing angle $\theta$ can thus be obtained by
\beq
\cos^{2} \theta=\frac{4m_{K_{1B}}^{2}-m_{b_{1}}^{2}-3m_{h_{1}(1170)}^{2}}{3\left(m_{h_{1}(1415)}^{2}-m_{h_{1}(1170)}^{2}\right)} \;, \qquad
\tan \theta=\frac{4m_{K_{1B}}^{2}-m_{b_{1}}^{2}-3m_{h_{1}(1415)}^{2}}{2\sqrt{2}\left(m_{b_{1}}^{2}-m_{K_{1B}}^{2}\right)} \;.
\label{eq:angle-thetaK2}
\eeq
The PDG results for the masses of physical states are used in our evaluation. It can be found that,
for a given $\theta_{K}$, the corresponding value of mixing angle $\theta$ can be extracted via the
above formulae, and the value of $\alpha$ can also be obtained by using Eq.~\eqref{eq:angle-relation}.
Using some possible values of $\theta_{K}$ as inputs, we give the results of $\theta$ and $\alpha$ in
Table~\ref{tab:two-mixing angles}. The default value $\theta=29.5^{\circ}$ used in our following 
calculations corresponds to $\theta_{K}=39^{\circ}$.
As has been shown in Eq.~\eqref{eq:mix-SO-QF}, the mixing angle $\theta$ plays an important
role in the investigation of $h_{1}$ mesons, thus the effects of $\theta$ on our theoretical
predictions are also discussed in our following analyses.

\begin{table}[htb]
\renewcommand{\arraystretch}{1.20}	
\caption{ The values of mixing angles of $h_{1}$ mesons in the QF (upper) and
    	 SO (lower) basis with some possible values of $\theta_{K}$ as inputs.}
\label{tab:two-mixing angles}
\begin{center}\vspace{-0.5cm}
\setlength{\tabcolsep}{4mm}{	
\begin{tabular}[t]{c|cccccc}
\hline \hline
\makebox[0.10\textwidth][c]{$|\theta_{K}|$ }
&\makebox[0.05\textwidth][c]{$27^{\circ}$ }
&\makebox[0.05\textwidth][c]{$33^{\circ}$ }
&\makebox[0.05\textwidth][c]{$39^{\circ}$ }
&\makebox[0.05\textwidth][c]{$45^{\circ}$ }
&\makebox[0.05\textwidth][c]{$51^{\circ}$ }
&\makebox[0.05\textwidth][c]{$57^{\circ}$ }
\\ \hline \hline
 $\alpha-90^{\circ}$
&$3.2^{\circ}$
&$-0.8^{\circ}$
&$-5.8^{\circ}$
&$-10.9^{\circ}$
&$-17.5^{\circ}$
&$-27.1^{\circ}$
\\ \hline
 $\theta$
&$38.5^{\circ}$
&$34.5^{\circ}$
&$29.5^{\circ}$
&$24.4^{\circ}$
&$17.8^{\circ}$
&$8.2^{\circ}$
\\ \hline \hline
\end{tabular}}
\end{center}
\end{table}

\subsection{ The \boldmath{$B^{0}\to J/\psi h_{1}$} decays in PQCD approach}
\label{sec0202}

The PQCD approach is one of the popular factorization approaches on the basis of QCD,
and has been widely employed to study a variety of $B$-meson decays. Recently, the
NLO PQCD predictions of the {\it CP}-averaged branching ratios for
the $B^{0}$-meson decays into a charmonium plus light hadrons have been improved
through including the important vertex corrections~\cite{Liu:2014doa,Liu:2019ymi,Yao:2022zom,Xiao:2019mpm,Ren:2023ebq}.
It makes this work a possible reference for future experimental measurements
and may provide a solid theoretical basis for exploring the possible
new physics beyond the SM potentially.

For the considered $B^0 \to J/\psi h_1$ decays, the effective Hamiltonian
can be written as~\cite{Buchalla:1995vs}
\beq
H_{\rm eff}\, &=&\, \frac{G_F}{\sqrt{2}}
\left\{ V^*_{cb}V_{cq} \biggl[ C_1(\mu)O_1^{c}(\mu)
+C_2(\mu)O_2^{c}(\mu) \biggr]
- V^*_{tb}V_{tq} \biggl[ \sum_{i=3}^{10}C_i(\mu)O_i(\mu) \biggr] \right\}\;,
\label{eq:heff}
\eeq
where $q=d$ or $s$, the Fermi constant $G_F=1.16639\times 10^{-5}{\rm GeV}^{-2}$,
$V$ stands for the Cabibbo-Kobayashi-Maskawa (CKM) matrix elements,
and $C_i(\mu)$ are Wilson coefficients at the renormalization scale $\mu$.
The local four-quark operators $O_i(i=1,\cdots,10)$ are given as
\begin{enumerate}
\item[]{(1) Tree operators}
\begin{eqnarray}
{\renewcommand\arraystretch{1.5}
\begin{array}{ll}
\displaystyle
O_1^{c}\, =\,(\bar{q}_\alpha c_\beta)_{V-A}(\bar{c}_\beta b_\alpha)_{V-A}\;,
& \displaystyle
O_2^{c}\, =\, (\bar{q}_\alpha c_\alpha)_{V-A}(\bar{c}_\beta b_\beta)_{V-A}\;,
\end{array}}
\label{eq:operators-1}
\end{eqnarray}
  	
\item[]{(2) QCD penguin operators}
\begin{eqnarray}
{\renewcommand\arraystretch{1.5}
\begin{array}{ll}
\displaystyle
O_3\, =\, (\bar{q}_\alpha b_\alpha)_{V-A}\sum_{q'}(\bar{q}'_\beta q'_\beta)_{V-A}\;,
& \displaystyle
O_4\, =\, (\bar{q}_\alpha b_\beta)_{V-A}\sum_{q'}(\bar{q}'_\beta q'_\alpha)_{V-A}\;,
\\
\displaystyle
O_5\, =\, (\bar{q}_\alpha b_\alpha)_{V-A}\sum_{q'}(\bar{q}'_\beta q'_\beta)_{V+A}\;,
& \displaystyle
O_6\, =\, (\bar{q}_\alpha b_\beta)_{V-A}\sum_{q'}(\bar{q}'_\beta q'_\alpha)_{V+A}\;,
\end{array}}
\label{eq:operators-2}
\end{eqnarray}
  	
\item[]{(3) Electroweak penguin operators}
\begin{eqnarray}
{\renewcommand\arraystretch{1.5}
\begin{array}{ll}
\displaystyle
O_7\, =\,
\frac{3}{2}(\bar{q}_\alpha b_\alpha)_{V-A}\sum_{q'}e_{q'}(\bar{q}'_\beta q'_\beta)_{V+A}\;,
& \displaystyle
O_8\, =\,
\frac{3}{2}(\bar{q}_\alpha b_\beta)_{V-A}\sum_{q'}e_{q'}(\bar{q}'_\beta q'_\alpha)_{V+A}\;,
\\
\displaystyle
O_9\, =\,
\frac{3}{2}(\bar{q}_\alpha b_\alpha)_{V-A}\sum_{q'}e_{q'}(\bar{q}'_\beta q'_\beta)_{V-A}\;,
& \displaystyle
O_{10}\, =\,
\frac{3}{2}(\bar{q}_\alpha b_\beta)_{V-A}\sum_{q'}e_{q'}(\bar{q}'_\beta q'_\alpha)_{V-A}\;,
\end{array}}
\label{eq:operators-3}
\end{eqnarray}
\end{enumerate}
with the color indices $\alpha, \beta$ and the notations
$V\pm A=  \gamma_\mu (1\pm \gamma_5)$.
The index $q'$ in the summation of the above operators runs
through $u,\;d,\;s$, $c$ and $b$. For convenience,
the combination $a_{i}$ of
Wilson coefficients are defined as~\cite{Liu:2013nea,Ali:2007ff}
\beq
a_{1}&=&C_{2}+\frac{C_{1}}{3}\;,
\qquad
a_{2}=C_{1}+\frac{C_{2}}{3}\;,
\eeq
and
\beq
a_{i} &=& C_{i}+\frac{C_{i\pm1}}{3}\quad(i=3-10)\;,
\label{eq:wilson}
\eeq
in which the upper(lower) sign applies, when $i$ is odd(even).

\begin{figure}[htp]
\centering
\includegraphics[scale=0.95]{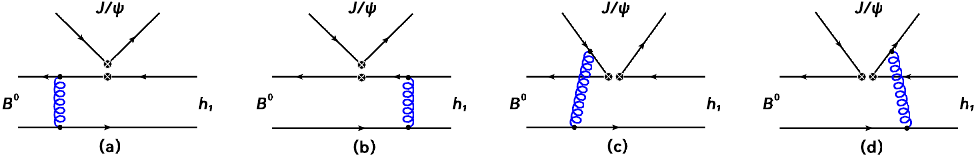}
\caption{ (Color online) Leading order Feynman diagrams for $B^0 \to J/\psi h_{1}$ in the PQCD approach.
	}
\label{fig:fig2}
\end{figure}

In Fig.~\ref{fig:fig2}, we show the typical Feynman diagrams contributing to
the $B^{0} \to J/\psi h_{1}$ decays in the PQCD approach at leading order (LO),
with the first two factorizable emission diagrams and the last two
non-factorizable emission ones. Consequently, analogous to the decays
$B^{0} \to J/\psi V$~\cite{Liu:2013nea}, the specific forms of the factorizable emission
amplitude $F_{J/\psi}^{h}$ ($h$ stands for three polarizations of the final (axial-)
vector states: longitudinal ($L$), normal ($N$), and transverse ($T$), respectively.)
and the non-factorizable emission amplitude $M_{J/\psi}^h$ are given as
\beq
F_{J/\psi}^{L}&=& 8\pi C_{F}f_{J/\psi}m_{B^{0}}^{4} \int_{0}^{1}dx_{1}dx_{3}
  \int_{0}^{\infty}b_{1}db_{1}b_{3}db_{3}\phi _{B}(x_{1},b_{1})\biggl\{\Bigl[ \sqrt{1-r_{2}^{2}} \bigl(r_{A} \bigl( \sqrt{1-r_{2}^{2}} \non
&& \times(2x_{3}-1)\phi_{A}^{s}(x_{3})-\phi_{A}^{t}(x_{3})(2x_{3}(r_{2}^{2}-1)+r_{2}^{2}+1)\bigr)
 +\phi_{A}(x_{3})\bigl((r_{2}^{2}-1)x_{3}-1\bigr) \bigr)\Bigr] \non
&& \times  h_{fe}(x_{1},x_{3},b_{1},b_{3})E_{fe}(t_{a})
 - \Bigl[2r_{A}(1-r_{2}^{2})
  \phi_{A}^{s}(x_{3})\Bigr]
  h_{fe}(x_{3},x_{1},b_{3},b_{1})E_{fe}(t_{b})\biggr\}\;,
\label{eq:fe-L}
\eeq
\beq
F_{J/\psi}^{N}&=& 8\pi C_{F}f_{J/\psi}m_{B^{0}}^{4} \int_{0}^{1}dx_{1}dx_{3}
  \int_{0}^{\infty}b_{1}db_{1}b_{3}db_{3}\phi_{B}(x_{1},b_{1})r_{2}
  \biggl\{\Bigl[(r_{2}^{2}-1)\bigl[r_{A}(r_{2}^{2}-1)x_{3} \non
&& \times \phi_{A}^{a}(x_{3})+\phi_{A}^{T}(x_{3})\bigr]
+r_{A}\bigl[(r_{2}^{2}-1)x_{3}-2\bigr]\phi_{A}^{v}(x_{3})\Bigr]
  h_{fe}(x_{1},x_{3},b_{1},b_{3})E_{fe}(t_{a}) \non
&&-r_{A}(r_{2}^{2}-1)\Bigl[(r_{2}^{2}-1)\phi_{A}^{a}
  (x_{3})-\phi_{A}^{v}(x_{3})\Bigr]h_{fe}(x_{3},x_{1},b_{3},b_{1})E_{fe}(t_{b}) \biggr\}\;,
\label{eq:fe-N}
\eeq
\beq
F_{J/\psi}^{T}&=& 16\pi C_{F}f_{J/\psi}m_{B^{0}}^{4} \int_{0}^{1}dx_{1}dx_{3}
  \int_{0}^{\infty}b_{1}db_{1}b_{3}db_{3}\phi _{B}(x_{1},b_{1})r_{2}\non
&& \times \biggl\{\Bigl[r_{A}x_{3}\phi_{A}^{v}(x_{3})
  -\phi_{A}^{T}(x_{3})+r_{A}\bigl[(r_{2}^{2}-1)x_{3}-2\bigr] \phi_{A}^{a}(x_{3})\Bigr]
  h_{fe}(x_{1},x_{3},b_{1},b_{3}) E_{fe}(t_{a})\non
&& +r_{A}\Bigl[(r_{2}^{2}-1)\phi_{A}^{a}(x_{3})-
  \phi_{A}^{v}(x_{3})\Bigr]
  h_{fe}(x_{3},x_{1},b_{3},b_{1})E_{fe}(t_{b})\biggr\}\;,
\label{eq:fe-T}
\eeq
with $r_{A}=m_{\tilde{h}_{1(8)}}/m_{B^{0}}$ and $r_{2}=m_{J/\psi}/m_{B^{0}}$, and
\beq
M_{J/\psi}^{L}&=& \frac{16\sqrt{6}}{3}\pi C_{F}m_{B^{0}}^{4}
  \int_{0}^{1}dx_{1}dx_{2}dx_{3}\int_{0}^{\infty}b_{1}db_{1}b_{2}db_{2}\phi_{B}(x_{1},b_{1})\non
&& \times \biggl\{\Bigl[ \sqrt{1-r_{2}^{2}} \bigl(\phi_{A}(x_{3})-2r_{A}
  \phi_{A}^{t}(x_{3})\bigr)\bigl(r_{2}^{2}\phi_{J/\psi}^{L}(x_{2})(2x_{2}-x_{3}) \non
&& +x_{3}\phi_{J/\psi}^{L}(x_{2})-2r_{2}r_{c}\phi_{J/\psi}^{t}(x_{2})\bigr)\Bigr]\biggr\}
  h_{nfe}(x_{1},x_{2},x_{3},b_{1},b_{2})E_{nfe}(t_{nfe})\;,
\label{eq:nfe-L}
\eeq
\beq
M_{J/\psi}^{N}&=& \frac{32\sqrt{6}}{3}\pi C_{F}m_{B^{0}}^{4}
  \int_{0}^{1}dx_{1}dx_{2}dx_{3}\int_{0}^{\infty}b_{1}db_{1}b_{2}db_{2} \phi_{B}(x_{1},b_{1})\non
&& \times \biggl\{(r_{2}^{2}-1)\Bigl[r_{2} x_{2}\phi_{J/\psi}^{v}(x_{2})
  -r_{c}\phi_{J/\psi}^{T}(x_{2})\Bigr]
  \phi_{A}^{T}(x_{3})+r_{A}\Bigl[-r_{c}(1+r_{2}^{2})\phi_{J/\psi}^{T}(x_{2}) \non
&& +r_{2}\bigl[x_{2}(1+r_{2}^{2})+x_{3}(1-r_{2}^{2})\bigr]
  \phi_{J/\psi}^{v}(x_{2})\Bigr]\phi_{A}^{v}(x_{3})\biggr\}
  h_{nfe}(x_{1},x_{2},x_{3},b_{1},b_{2})E_{nfe}(t_{nfe})\;,
\label{eq:nfe-N}
\eeq
\beq
M_{J/\psi}^{T}&=& \frac{64\sqrt{6}}{3}\pi C_{F}m_{B^{0}}^{4}
  \int_{0}^{1}dx_{1}dx_{2}dx_{3}\int_{0}^{\infty}b_{1}db_{1}b_{2}db_{2}\phi_{B}(x_{1},b_{1})\non
&& \times \biggl\{\Bigl[r_{c}\phi_{J/\psi}^{T}(x_{2})-r_{2}x_{2}\phi_{J/\psi}^{v}(x_{2})\Bigr]
  \phi_{A}^{T}(x_{3})+r_{A}\Bigl[-r_{c}(1+r_{2}^{2})\phi_{J/\psi}^{T}(x_{2})+r_{2}
  \bigl[x_{2}(1+r_{2}^{2}) \non
&&+x_{3}(1-r_{2}^{2})\bigr] \phi_{J/\psi}^{v}(x_{2})\Bigr]\phi_{A}^{a}(x_{3})\biggr\}
  h_{nfe}(x_{1},x_{2},x_{3},b_{1},b_{2})E_{nfe}(t_{nfe})\;,
  \label{eq:nfe-T}
  \eeq
where $r_{c}=m_{c}/m_{B^{0}}$ with $m_{c}$ being the charm quark mass.
For the sake of simplicity, the explicit forms of hard function $h(x_{i},b_{i})$,
evolution function $E_{i}(t)$ and the running hard scale $t$
of $F_{J/\psi}^{h}$ and $M_{J/\psi}^{h}$ in the above equations
\eqref{eq:fe-L}-\eqref{eq:nfe-T} can be referred to the Refs.~\cite{Liu:2013nea,Ren:2023ebq}.

\begin{figure}[!!htp]
	\centering
	\includegraphics[scale=1.0]{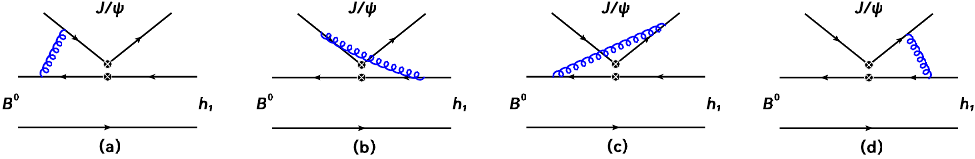}
	\caption{(Color online) Vertex corrections to $B^0 \to J/\psi h_{1}$ at NLO }
	\label{fig:fig3}
\end{figure}

As has been emphasized in the introduction, the color-suppressed $B^{0} \to J/\psi h_{1}$
decays should include the known NLO contributions from vertex corrections and NLO Wilson
coefficients to improve the accuracy of the theoretical predictions.
The corresponding vertex corrections have been shown in Fig.~\ref{fig:fig3}, and will be considered
in our calculation. As has been pointed out in~\cite{Chen:2005ht,Cheng:2000kt}, their effects
can be absorbed into the Wilson coefficients of factorizable contributions
and subsequently form a set of effective Wilson coefficients $\tilde{a}_{i}^{\sigma}(i=2,3,5,7,9)$ with helicities $(\sigma=0,\pm)$. 
The explicit expressions of $\tilde{a}_{i}^{\sigma}$
can be found in Ref.~\cite{Yao:2022zom}.

In the PQCD calculations at LO accuracy, we shall use the LO Wilson coefficients
$C_{i}(m_{W})$, the LO renormalization group (RG) evolution matrix $U(t,m)^{(0)}$
for the Wilson coefficient with the LO running coupling $\alpha_{s}(t)^{(0)}$,
\beq
\alpha_{s}(t)^{(0)}=\frac{4\pi}{\beta_{0}\ln\left(t^{2}/\Lambda_{\rm QCD}^{2}\right)}\;,
\eeq
where $\beta_{0}=(33-2N_{f})/3$. While, the NLO Wilson coefficients $C_{i}(m_{W})$
and the NLO RG evolution matrix $U(t,m,\alpha)$ with the running coupling $\alpha_{s}(t)$
at two-loop level should be used in the PQCD calculations at the NLO accuracy~\cite{Buchalla:1995vs},
\beq
\alpha_{s}(t)=\frac{4\pi}{\beta_{0}\ln\left(t^{2}/\Lambda_{\rm QCD}^{2}\right)}\, \Biggl
\{1-\frac{\beta_{1}}
{\beta_{0}^{2}}\cdot \frac{\ln
	\left[\ln\left(t^{2}/\Lambda_{\rm QCD}^{2}\right)\right]}{\ln\left(t^{2}/\Lambda_{\rm QCD}^{2}\right)}\Biggr\}\;,
\eeq
where $\beta_{0}=(33-2N_{f})/3$ and $\beta_{1}=(306-38N_{f})/3$. The hadronic scale
$\Lambda_{\rm QCD}^{(4)}=0.287$ GeV (0.326 GeV)
could be obtained by using $\Lambda_{\rm QCD}^{(5)}=0.225$ GeV for the LO (NLO)
case~\cite{Buchalla:1995vs}. For the hard scale $t$,
the lower cut-off $\mu_{0}=1.0$ GeV is chosen~\cite{Xiao:2008sw}.

Using the building-blocks given above, the decay amplitudes of $B^0 \to J/\psi \tilde{h}_{1(8)}$
can then be written as
\beq
A^{h}(B_{d,s}^0 \to J/\psi \tilde{h}_{1(8)})&=&
F_{J/\psi}^{h} \Biggl\{V_{cb}^{\ast} V_{cd(s)} \tilde{a}_{2}^{\sigma}-V_{tb}^{\ast} V_{td(s)} \biggl(\tilde{a}_{3}^{\sigma}+\tilde{a}_{5}^{\sigma}
+\tilde{a}_{7}^{\sigma}+\tilde{a}_{9}^{\sigma}\biggr)  \Biggr\} \nonumber\\
&&+ M_{J/\psi}^{h}\Biggl\{V_{cb}^{\ast} V_{cd(s)}C_{2}-V_{tb}^{\ast} V_{td(s)} \biggl(C_{4}-C_{6}-C_{8}+C_{10}\biggr)\Biggr\} \;,
\label{eq:DecAmp-jpsi}
\eeq
with the superscripts $h$ and $\sigma$ representing the polarization and helicity of the final states, respectively.
Specifically, $h=L$ corresponds to a helicity $\sigma=0$, while $h=N,T$ corresponds to helicities $\sigma=\pm$.
Combining various contributions from different Feynman diagrams and the single-octet mixing scheme
as given in Eq.~\eqref{eq:mix-SO-QF}, the decay amplitudes of the considered $B^0 \to J/\psi h_{1}$
decays with physical states could be expressed as follows,
\begin{itemize}
\item[(1)]{For $B_d^0 \to J/\psi h_{1} $ decays,}
\beq
{\cal A}^{h}\left(B_{d}^{0}\to J/\psi h_{1}(1170)\right)
&=& \; A^{h}\left(B_{d}^{0}\to J/\psi \tilde{h}_{1}\right)\frac{\cos \theta}{\sqrt{3}} + A^{h}\left(B_{d}^{0}\to J/\psi \tilde{h}_{8}\right)
\frac{\sin \theta}{\sqrt{6}}\;,
\label{eq:DecAmp-psih11-d}
\\
{\cal A}^{h}\left(B_{d}^{0}\to J/\psi h_{1}(1415)\right)
&=& \; A^{h}\left(B_{d}^{0}\to J/\psi \tilde{h}_{8}\right)\frac{\cos \theta}{\sqrt{6}} - A^{h}\left(B_{d}^{0}\to J/\psi \tilde{h}_{1}\right)
\frac{\sin \theta}{\sqrt{3}}\;,
\label{eq:DecAmp-psih14-d}
\eeq
  	
\item[(2)]{For $B_s^0 \to J/\psi h_{1} $ decays,}
\beq
{\cal A}^{h}\left(B_{s}^{0}\to J/\psi h_{1}(1170)\right)
&=& \;A^{h}\left(B_{s}^{0}\to J/\psi \tilde{h}_{1}\right)\frac{\cos \theta}{\sqrt{3}} - 2A^{h}\left(B_{s}^{0}\to J/\psi \tilde{h}_{8}\right)
\frac{\sin \theta}{\sqrt{6}}\;,
\label{eq:DecAmp-psih11-s}
\\
{\cal A}^{h}\left(B_{s}^{0}\to J/\psi h_{1}(1415)\right)
&=&-2A^{h}\left(B_{s}^{0}\to J/\psi \tilde{h}_{8}\right) \frac{\cos \theta}{\sqrt{6}} - A^{h}\left(B_{s}^{0}\to J/\psi \tilde{h}_{1}\right)
\frac{\sin \theta}{\sqrt{3}}\;.\;\;
\label{eq:DecAmp-psih14-s}
\eeq
\end{itemize}

\section{Numerical results and discussions}
\label{sec03}

\begin{table}[htb]
\renewcommand{\arraystretch}{1.30}
\caption{The values of input parameters.}
\label{tab:inputs}
\begin{center}\vspace{-0.5cm}
\begin{tabular}[t]{l|lll}
\hline
\hline
\multirow{2}{*}{Masses (GeV)}  &$m_{B_{d}^{0}}=5.28$\,,\quad$m_{B_s^0}=5.37$\,,\quad$m_{b}=4.8$\,,\quad$m_{c}=1.50$\,,\quad$m_{J/\psi}=3.097$\;,\\
& $m_{h_{1}(1170)}=1.166$\,,\quad$m_{h_{1}(1415)}=1.409$\;, \quad$m_{\tilde{h}_{1}}=1.23$\;, \quad$m_{\tilde{h}_{8}}=1.36$\\
\hline
\multirow{2}{*}{Decay constants (GeV)}
&$f_{B_{d}^{0}}=0.21\pm0.02$\,,\quad$f_{B_{s}^{0}}=0.23\pm0.02$\,,\quad$f_{J/\psi}=0.405\pm0.014$\,,\\
&$f_{\tilde{h}_{1}}=0.180\pm0.012$\,,\quad$f_{\tilde{h}_{8}}=0.190\pm0.010$\\
\hline
$B$-meson lifetimes (ps)
&$\tau_{B_d^0}=1.519\pm0.004$\,,\quad$\tau_{B_s^0}= 1.520\pm0.005$ \\
\hline
\multirow{2}{*}{CKM
parameters}  \qquad \qquad
&$A=0.826^{+0.018}_{-0.015}$\,,\quad  $\lambda=0.22500\pm0.00067$\,, \quad $\bar{\rho}=0.159\pm0.010$\,, \\
&$\bar{\eta}=0.348\pm0.010$\\
\hline
\hline 
\end{tabular}
\end{center}\vspace{-0.5cm}
\end{table}

In the numerical calculations, the meson masses, decay constants, $B^{0}$ meson lifetimes, and
CKM matrix elements~(Wolfenstein parameters~\cite{Wolfenstein:1983yz} ) are essential input
parameters. Their values~\cite{ParticleDataGroup:2022pth,Yang:2007zt,Verma:2011yw} are collected
in Table~\ref{tab:inputs}. For the nonleptonic two-body $B^0 \to J/\psi h_{1}$ decays,
the branching fraction ${\cal B}$ is defined as
\beq
{\cal B} &\equiv& \tau_{B^{0}}\cdot \Gamma\left(B^0 \to J/\psi h_{1}\right)
= \tau_{B^{0}}\cdot \frac{G_{F}^{2}\left|\bf{P_c}\right|}{16 \pi m^{2}_{B^{0}} }
\sum_{h=L,N,T} {\cal A}^{(h)\dagger } {\cal A}^{(h)}\;,
\label{eq:br-psih1}
\eeq
where $\tau_{B^{0}}$ is the lifetime of $B^{0}$-meson, $|\bf{P_c}|\equiv |\bf{P_{2z}}|=|\bf{P_{3z}}|$
is the three-momentum of outgoing final states, and ${\cal A}^h$ denotes the helicity amplitudes of
$B^0 \to J/\psi h_{1}$ decays given in Eqs.~\eqref{eq:DecAmp-psih11-d}-\eqref{eq:DecAmp-psih14-s}.

Using the decay amplitudes and input parameters given above, our LO and NLO PQCD predictions
for the {\it CP}-averaged branching fractions of $B^0 \to J/\psi h_{1}$ decays, accompanied with
multiple uncertainties, are given in Table~\ref{tab:branching ratio}. The first four theoretical
errors are induced by the shape parameter $\omega_{B_d^0} = 0.40 \pm 0.04$~GeV or
$\omega_{B_s^0} = 0.50 \pm 0.05$~GeV in the $B^{0}$-meson distribution amplitude,
the decay constant $f_{M}$ of two outgoing final states as presented in Table~\ref{tab:inputs},
the charm quark mass $m_{c}=1.50\pm0.15$~GeV, and the Gegenbauer moments $a_{1}^{\parallel}$ and
$a_{2}^{\perp}$ (see Eq.~\eqref{eq:Geng-h1}) in the light-cone distribution amplitudes of
the $h_{1}$ mesons, respectively. The last error arises from factor $a_{t}=1.0 \pm 0.2$
describing the possible higher-order corrections, which are characterized through simply varying
the running hard scale $t_{\rm max}$ with $20\%$ in the hard kernel.
From Table~\ref{tab:branching ratio}, it can be clearly seen that the dominated uncertainties
arise mainly from the hadronic parameters such as the Gegenbauer moments and the decay constants.

\begin{table}[htb]
\renewcommand{\arraystretch}{1.25}	
\caption{ The LO and NLO PQCD predictions of the {\it CP}-averaged branching ratios for the $B^{0} \to J/\psi h_{1}$ decays. }
\label{tab:branching ratio}
\begin{center}\vspace{-0.5cm}	
\begin{tabular}[t]{c|c|c}
\hline  \hline
\makebox[0.20\textwidth][c]{Decay modes}
& \makebox[0.30\textwidth][c]{LO}
& \makebox[0.30\textwidth][c]{NLO}  \\
\hline \hline
$B_{d}^{0} \to J/\psi h_{1}(1170)$
& $\begin{array}{cc} \left(6.55_{-1.35-0.91-0.90-1.31-1.60}^{+1.78+0.96+1.20+1.58+2.23}\right) \times 10^{-5} \end{array} $
& $\begin{array}{cc} \left(9.54_{-1.87-1.54-0.74-2.43-0.38}^{+2.39+1.65+0.72+2.83+0.26}\right) \times 10^{-5} \end{array} $
\\ \hline
$B_{d}^{0} \to J/\psi h_{1}(1415)$
& $\begin{array}{cc} \left(1.12_{-0.26-0.08-0.08-0.17-0.32}^{+0.36+0.08+0.14+0.22+0.44}\right) \times 10^{-6} \end{array} $
& $\begin{array}{cc} \left(1.32_{-0.26-0.15-0.13-0.35-0.07}^{+0.35+0.17+0.14+0.41+0.04}\right) \times 10^{-6} \end{array} $
\\ \hline
$B_{s}^{0} \to J/\psi h_{1}(1170)$
& $\begin{array}{cc} \left(0.90_{-0.20-0.23-0.03-0.22-0.05}^{+0.27+0.26+0.09+0.26+0.08}\right) \times 10^{-5} \end{array} $
& $\begin{array}{cc} \left(1.50_{-0.32-0.39-0.03-0.38-0.02}^{+0.42+0.44+0.01+0.44+0.01}\right) \times 10^{-5} \end{array} $
\\ \hline
$B_{s}^{0} \to J/\psi h_{1}(1415)$
& $\begin{array}{cc} \left(1.05_{-0.23-0.14-0.06-0.30-0.04}^{+0.29+0.14+0.09+0.36+0.07}\right) \times 10^{-3} \end{array} $
& $\begin{array}{cc} \left(1.74_{-0.39-0.22-0.02-0.51-0.02}^{+0.50+0.26+0.01+0.61+0.01}\right) \times 10^{-3} \end{array} $
\\ \hline \hline
\end{tabular}
\end{center}
\end{table}

Comparing the numerical results at LO with the ones at NLO, we find that the NLO contributions
from vertex corrections can significantly enhance the branching ratios due to the corrected
effective Wilson coefficient $\tilde{a}_{2}$ being much larger than the original one $a_{2}$~\cite{Cheng:2000kt,Chen:2005ht}.
Furthermore, we also note that the error arisen from the running hard scale $t$ can be reduced from $30\%$ to $5\%$.
Therefore, the subsequent numerical results and phenomenological analyses are all at the presently known NLO level,
unless otherwise specified.

Adding the above-mentioned errors in quadrature, the NLO  PQCD results for ${\cal B}(B^{0} \to J/\psi h_{1})$ can be written as
\begin{itemize}
\item[(i)]{For $\bar{b} \to c\bar{c} \bar{d} $ decays,}
\beq
{\cal B}\left(B_{d}^{0}\to J/\psi h_{1}(1170)\right)=9.54_{-3.53}^{+4.13} \times 10^{-5}, \;\;
\label{eq:BR-psih11-d}
{\cal B}\left(B_{d}^{0}\to J/\psi h_{1}(1415)\right)=1.32_{-0.48}^{+0.58} \times 10^{-6},\;\;
\label{eq:BR-psih14-d}
\eeq
	
\item[(ii)]{For $\bar{b} \to c\bar{c} \bar{s} $ decays,}
\beq
{\cal B}\left(B_{s}^{0}\to J/\psi h_{1}(1170)\right)=1.50_{-0.63}^{+0.75} \times 10^{-5}, \;\;
\label{eq:BR-psih11-s}
{\cal B}\left(B_{s}^{0}\to J/\psi h_{1}(1415)\right)=1.74_{-0.68}^{+0.83} \times 10^{-3}. \;\;
\label{eq:BR-psih14-s}
\eeq
\end{itemize}
From these results, it can be found that $B^{0}\to J/\psi h_{1}$ decays have relatively large
branching ratios, which are around $10^{-6} \sim 10^{-3}$, and are possible to be measured
at the ongoing LHCb and Belle-II experiments in the near future.

\begin{table}[ht]
\renewcommand{\arraystretch}{1.10}	
\caption{ Decay amplitudes (in units of $10^{-3} \;{\rm GeV}^{3})$ of the $B^{0} \to J/\psi h_{1}$ modes
with three polarizations in the PQCD approach.
}
\label{tab:three amp-phy}
\begin{center}\vspace{-0.5cm}	
\begin{tabular}[t]{c|c||r|r|r|r}
\hline  \hline
\multicolumn{2}{c||}{Decays}
&\multicolumn{2}{c|}{${ \cal A}_{B_{d}^{0}}$}
&\multicolumn{2}{c}{${ \cal A}_{B_{s}^{0}}$}
\\  \hline
\multicolumn{2}{c||}{Contributions}
&\makebox[0.16\textwidth][c]{Tree}
&\makebox[0.16\textwidth][c]{Penguin}
&\makebox[0.16\textwidth][c]{Tree}
&\makebox[0.16\textwidth][c]{Penguin}
\\ \hline \hline
\makebox[0.10\textwidth][c]{$\;$}
&\makebox[0.08\textwidth][c]{${\cal A}_{L}$}
&$ -0.125-i0.529  $
&$ -0.002-i0.013  $
&$ 0.026+i0.212  $
&$ -0.001+i0.006  $
\\
$J/\psi h_{1}(1170)$
&${\cal A}_{N}$
&$ -0.032-i0.143  $
&$ 0.000-i0.004  $
&$ 0.026+i0.034  $
&$ 0.000+i0.001  $
\\
$\;$
&${\cal A}_{T}$
&$ -0.110-i0.394  $
&$ -0.001-i0.010  $
&$ 0.080+i0.104  $
&$ 0.001+i0.003  $
\\ \hline
$\;$
&${\cal A}_{L}$
&$ -0.021-i0.064  $
&$ -0.000-i0.002  $
&$ -0.296-i2.422  $
&$ 0.003-i0.061  $
\\
$J/\psi h_{1}(1415)$
&${\cal A}_{N}$
&$ -0.002-i0.020  $
&$ 0.000-i0.001  $
&$ -0.235-i0.410  $
&$ -0.005-i0.010  $
\\
$\;$
&${\cal A}_{T}$
&$ -0.008-i0.053  $
&$  0.000-i0.002  $
&$ -0.705-i1.227  $
&$ -0.014-i0.030  $
\\ \hline \hline
\end{tabular}
\end{center}\vspace{-0.5cm}
\end{table}

\begin{table}[ht]
\renewcommand{\arraystretch}{1.10}	
\caption{ Same as Table~\ref{tab:three amp-phy} but for neutral $B$-meson decays into $J/\psi \tilde{h}_{1}$
and $J/\psi \tilde{h}_{8}$. }
\label{tab:three amp-SO}
\begin{center}\vspace{-0.5cm}	
\begin{tabular}[t]{c|c||r|r|r|r}
\hline  \hline
\multicolumn{2}{c||}{Decays}
&\multicolumn{2}{c|}{$A_{B_{d}^{0}}$}
&\multicolumn{2}{c}{$A_{B_{s}^{0}}$}
\\  \hline
\multicolumn{2}{c||}{Contributions}
&\makebox[0.16\textwidth][c]{Tree}
&\makebox[0.16\textwidth][c]{Penguin}
&\makebox[0.16\textwidth][c]{Tree}
&\makebox[0.16\textwidth][c]{Penguin}
\\ \hline \hline
\makebox[0.10\textwidth][c]{$\;$}
&\makebox[0.08\textwidth][c]{${\cal A}_{L}$}
&$ -0.170-i0.743  $
&$ -0.003-i0.018  $
&$ 0.292+i2.384  $
&$ -0.003+i0.060 $
\\
$J/\psi \tilde{h}_{1}$
&${\cal A}_{N}$
&$ -0.047-i0.198  $
&$ 0.000-i0.005  $
&$ 0.239+i0.401  $
&$ 0.005+i0.010  $
\\
$\;$
&${\cal A}_{T}$
&$ -0.159-i0.549  $
&$ -0.001-i0.014  $
&$ 0.721+i1.204  $
&$ 0.014+i0.029  $
\\ \hline
$\;$
&${\cal A}_{L}$
&$ -0.195-i0.775  $
&$ -0.003-i0.019  $
&$ 0.300+i2.454  $
&$ -0.003+i0.061 $
\\
$J/\psi \tilde{h}_{8}$
&${\cal A}_{N}$
&$ -0.043-i0.214  $
&$ 0.000-i0.006  $
&$ 0.234+i0.401  $
&$ 0.005+i0.010  $
\\
$\;$
&${\cal A}_{T}$
&$ -0.151-i0.549  $
&$ -0.001-i0.015  $
&$ 0.704+i1.245  $
&$ 0.014+i0.030  $
\\ \hline \hline
\end{tabular}
\end{center}\vspace{-0.5cm}
\end{table}

\begin{figure}[!!htb]
\begin{center}
\includegraphics[scale=0.60]{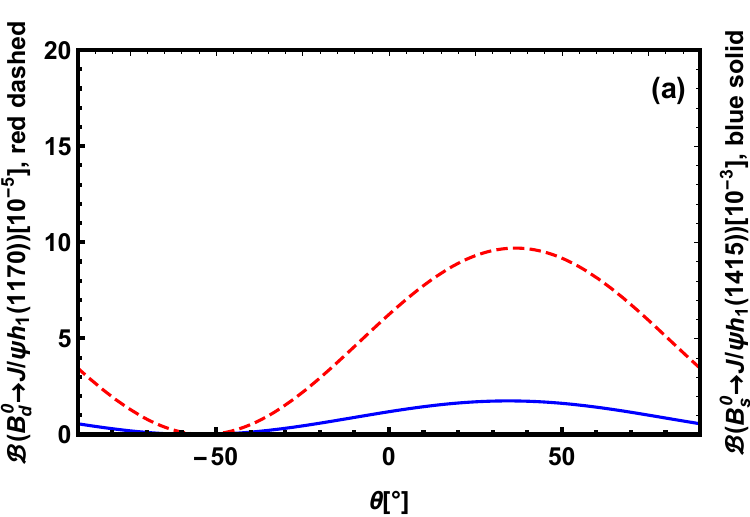}\hspace{1.0cm}
\includegraphics[scale=0.60]{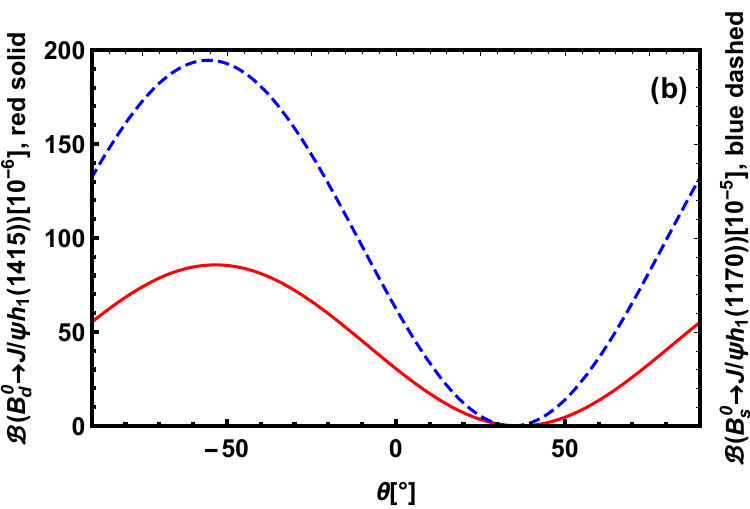}
\caption{(Color online) The dependence of ${\cal B}(B^{0} \to J/\psi h_{1})$ on the mixing angle $\theta$ in the PQCD approach.
 }
\label{fig:fig4}
\end{center}\vspace{-0.5cm}
\end{figure}

In order to see clearly the interferences between two final states $J/\psi \tilde{h}_{1}$
and $J/\psi \tilde{h}_{8}$, we also give the numerical results of polarization amplitudes
of $B^{0} \to J/\psi h_{1}$ and $B^{0} \to J/\psi \tilde{h}_{1,8}$ modes in
Tables~\ref{tab:three amp-phy} and \ref{tab:three amp-SO}, respectively.
Meanwhile, the amplitudes induced by the tree operators and the penguin operators
are also listed. These numerical results indicate that the considered decay modes
are dominated by the tree diagrams, with only a few percent of penguin contaminations.
Combined with Eqs.~\eqref{eq:DecAmp-psih11-d}-\eqref{eq:DecAmp-psih14-s}, it is evident
to observe that the constructive or destructive interferences with different extents
in the considered decays could vary with the mixing angle $\theta$ in the SO basis.
Thus, in order to show the effects of mixing angle $\theta$, we plot the dependence
of NLO PQCD predictions of ${\cal B}(B^{0} \to J/\psi h_{1})$ on the mixing angle
$\theta \in[-\frac{\pi}{2},\frac{\pi}{2}]$ in Fig.~\ref{fig:fig4}. It can be clearly
seen that the ${\cal B}(B^{0} \to J/\psi h_{1})$ are very sensitive to the mixing angle $\theta$.
From the Eqs.~\eqref{eq:DecAmp-psih11-d}-\eqref{eq:DecAmp-psih14-s} and  Fig.~\ref{fig:fig4},
it can be clearly seen that the interference between the $B_{s}^{0} \to J/\psi \tilde{h}_{1}$
and $B_{s}^{0} \to J/\psi \tilde{h}_{8}$ decay amplitudes consequently leads to
that ${\cal B} (B_{s}^{0}\to J/\psi h_{1}(1170), \,B_{d}^{0} \to J/\psi h_{1}(1415) )$
are significantly reduced, while ${\cal B} (B_{d}^{0}\to J/\psi h_{1}(1170), \,B_{s}^{0}
\to J/\psi h_{1}(1415) )$ are enhanced, at $\theta\sim 29.5^{\circ}$. Therefore,
the experimental measurement of these branching fractions would
play an important role in testing the value of $\theta$.

There are several unflavored light mesons, such as $a_{1}(1260),$ $f_{1}(1285),$ $\eta(1405),$
$f_{1}(1420)$ and $\eta(1475)$, that are known to decay into $K_S^{0}K^{\pm}\pi^{\mp}$, and that
could in principle be produced in $B^{0}$ meson decays alongside charmonium states~\cite{LHCb:2014sli}.
For the $h_{1}(1415)$ state below 1500 ${\rm MeV}/c^{2}$, the process $h_{1}(1415)\to K_S^{0}K^{\pm}\pi^{\mp}$
has not been measured~\cite{ParticleDataGroup:2022pth}.
As discussed in Ref.~\cite{Du:2021zdg}, the decay rate of the dominant decay mode is
${\cal B}(h_{1}(1415) \to {K}^{\ast} \bar{K})=0.415\pm0.085$, then the branching fraction
${\cal B}(h_{1}(1415) \to  K_S^{0}K^{\pm}\pi^{\mp}) \approx 0.277 \pm0.056$ can be naively
determined due to the isospin conservation for the strong decays $K^\ast \to K \pi$.
Based on the narrow-width-approximation (NWA), we can obtain the branching fractions of
$B^{0} \to J/\psi h_{1}(1415)(\to K_{S}^{0}K^{+}\pi^{-})$ decays,
\beq
{\cal B}\left(B_{s}^{0} \to J/\psi (K_{S}^{0}K^{+}\pi^{-})_{h_{1}(1415)}\right)&\equiv &
{\cal B}\left(B_{s}^{0} \to J/\psi h_{1}(1415)\right) \notag  \\
&&\cdot {\cal B}\left(h_{1}(1415) \to K_{S}^{0}K^{+}\pi^{-}\right)
\approx 0.48_{-0.21}^{+0.25}\times 10^{-3} \;,
\label{eq:br2-s}
\eeq
\beq
{\cal B}\left(B_{d}^{0} \to J/\psi (K_{S}^{0}K^{+}\pi^{-})_{h_{1}(1415)}\right)&\equiv &
{\cal B}\left(B_{d}^{0} \to J/\psi h_{1}(1415)\right)  \notag  \\
&&\cdot {\cal B}\left(h_{1}(1415) \to K_{S}^{0}K^{+}\pi^{-}\right)
\approx 0.37_{-0.15}^{+0.18}\times 10^{-6} \;,
\label{eq:br2-d}
\eeq
where the error arising from the $h_{1}(1415)$ decay width has been taken into account.
As have been given in Ref.~\cite{Yao:2022zom}, the theoretical results for above processes
through $f_{1}(1420)$ resonance showed that,
\beq
\label{eq:br2-theo-f1-s}
 {\cal B}(B_{s}^{0} \to J/\psi (K_{S}^{0}K^{+}\pi^{-})_{f_{1}(1420)}) \approx 0.73_{-0.28}^{+0.36}\times 10^{-3}\;,\\
 {\cal B}(B_{d}^{0} \to J/\psi (K_{S}^{0}K^{+}\pi^{-})_{f_{1}(1420)}) \approx 1.27_{-0.68}^{+0.88}\times 10^{-6} \;.
\label{eq:br2-theo-f1-d}
\eeq
While, the $B^{0}\to J/\psi K_{S}^{0}K^{\pm}\pi^{\mp}$ decays have been measured by the LHCb Collaboration,
their branching ratios are~\cite{LHCb:2014sli}
\beq
\label{eq:br2-exp-s}
{\cal B}(B_{s}^{0} \to J/\psi K_{S}^{0}K^{+}\pi^{-})_{\rm Exp}&=&(0.91 \pm 0.09)\times 10^{-3}\;,  \\
{\cal B}(B_{d}^{0} \to J/\psi K_{S}^{0}K^{+}\pi^{-})_{\rm Exp}&<&2.01\times 10^{-5}\;.
\label{eq:br2-exp-d}
\eeq
For the well measured $B_{s}^{0} \to J/\psi K_{S}^{0}K^{+}\pi^{-}$ decay, comparing the theoretical results given
in Eqs.~\eqref{eq:br2-s} and \eqref{eq:br2-theo-f1-s} with data given in Eq.~\eqref{eq:br2-exp-s}, it can be
found that the sum of the branching fractions of $B_{s}^{0} \to J/\psi K_{S}^{0}K^{+}\pi^{-}$ decay via
$h_{1}(1415)$ and $f_{1}(1420)$ resonances is approximately consistent with the LHCb measurement of ${\cal B}(B_{s}^{0} \to J/\psi K_{S}^{0}K^{+}\pi^{-})$. However, considering the presently unknown interferences between the amplitudes from two different $f_{1}(1420)$ and $h_{1}(1415)$ resonant states, an accurate quantification of the $h_{1}(1415)$ resonance contributing to the $m(K_{S}^{0}K^{\pm}\pi^{\mp})$ distribution still requires further experimental and/or theoretical examinations 
on the interference effects.

As has been shown in Table~\ref{tab:branching ratio}, the branching fractions of the $B^{0} \to J/\psi h_{1}$
decays have relatively large theoretical uncertainties due to the various hadronic input parameters.
Generally, the theoretical errors caused by the same hadronic input parameters can be significantly cancelled
by introducing some ratios. For instance, one can define the following two ratios,
\beq
R_{d}^{\rm SO} &\equiv& \frac{{\cal B}\left(B_{d}^{0} \to J/\psi h_{1}(1170)\right)}{{\cal B}\left(B_{d}^{0} \to J/\psi h_{1}(1415)\right)} \notag \\
&=& \frac{\Phi\left(m_{B_{d}^{0}},m_{J/\psi},m_{h_{1}(1170)}\right)}{\Phi\left(m_{B_{d}^{0}},m_{J/\psi},m_{h_{1}(1415)}\right)} \cdot
\frac{\left|\frac{\cos \theta}{\sqrt{3}} \cdot A(B_{d}^{0} \to J/\psi \tilde{h}_{1}) +
\frac{\sin \theta}{\sqrt{6}} \cdot A(B_{d}^{0} \to J/\psi \tilde{h}_{8})\right|^{2}}
{\left|-\frac{\sin \theta}{\sqrt{3}} \cdot A(B_{d}^{0} \to J/\psi \tilde{h}_{1}) +
\frac{\cos \theta}{\sqrt{6}} \cdot A(B_{d}^{0} \to J/\psi \tilde{h}_{8})\right|^{2}}\;,
\eeq
\beq
R_{s}^{\rm SO} &\equiv& \frac{{\cal B}\left(B_{s}^{0} \to J/\psi h_{1}(1415)\right)}{{\cal B}\left(B_{s}^{0} \to J/\psi h_{1}(1170)\right)} \notag \\
&=&\frac{\Phi\left(m_{B_{s}^{0}},m_{J/\psi},m_{h_{1}(1415)}\right)}{\Phi\left(m_{B_{s}^{0}},m_{J/\psi},m_{h_{1}(1170)}\right)} \cdot
\frac{\left|-\frac{\sin \theta}{\sqrt{3}} \cdot A(B_{s}^{0} \to J/\psi \tilde{h}_{1}) -2\cdot
\frac{\cos \theta}{\sqrt{6}} \cdot A(B_{s}^{0} \to J/\psi \tilde{h}_{8})\right|^{2}}
{\left|\frac{\cos \theta}{\sqrt{3}} \cdot A(B_{s}^{0} \to J/\psi \tilde{h}_{1}) -2\cdot
\frac{\sin \theta}{\sqrt{6}} \cdot A(B_{s}^{0} \to J/\psi \tilde{h}_{8})\right|^{2}},\;\;
\eeq
where the phase space factor is given as $\Phi(a,b,c)=[(a^{2}-(b+c)^{2})(a^{2}-(b-c)^{2})]^{\frac{1}{2}}$~\cite{Fleischer:2011au}.
It can be found that, these ratios can be used to test or extract the values of mixing angle $\theta$ approximately
in a model independent way if $A(B^{0} \to J/\psi \tilde{h}_{1}) = A(B^{0} \to J/\psi \tilde{h}_{8})$,
which is approximately valid as has been shown through the numerical results given in Table~\ref{tab:three amp-SO}.
Our numerical results are
\beq
R_{d}^{\rm SO}=72.27_{-4.53}^{+3.59}\;,
\qquad
R_{s}^{\rm SO}=116.52_{-10.96}^{+10.46}\;.
\eeq
It can be obviously found that the theoretical errors are significantly reduced.
Moreover, in order to obtain a more intuitive interpretation, one can employ a
more convenient and intuitive form by extending the ratios to the QF basis,
which can be exactly derived as
\beq
R_{d}^{\rm QF} \equiv \frac{{\cal B}\left(B_{d}^{0} \to J/\psi h_{1}(1170)\right)}{{\cal B}\left(B_{d}^{0} \to J/\psi h_{1}(1415)\right)}
= \frac{\Phi\left(m_{B_{d}},m_{J/\psi},m_{h_{1}(1170)}\right)}{\Phi\left(m_{B_{d}},m_{J/\psi},m_{h_{1}(1415)}\right)} \cdot \cot^{2} \alpha\;,
\eeq
\beq
R_{s}^{\rm QF} \equiv \frac{{\cal B}\left(B_{s}^{0} \to J/\psi h_{1}(1415)\right)}{{\cal B}\left(B_{s}^{0} \to J/\psi h_{1}(1170)\right)}
= \frac{\Phi\left(m_{B_{s}},m_{J/\psi},m_{h_{1}(1415)}\right)}{\Phi\left(m_{B_{s}},m_{J/\psi},m_{h_{1}(1170)}\right)} \cdot \cot^{2} \alpha\;,
\eeq
where the ratios $R_{d}^{\rm QF} $ and $R_{s}^{\rm QF} $ are independent on the decay amplitudes.
They could provide a much more convenient and clear way to extract the angle $\alpha$ when they
are measured by the future experiments. The mixing angle $\theta$ can be further obtained via
Eq.~\eqref{eq:angle-relation}. Beside of the ratios mentioned above, the ratios defined as
\beq
R_{ds}^{\rm SO}[h_{1}(1170)] \equiv \frac{{\cal B}\left(B_{d}^{0} \to J/\psi h_{1}(1170)\right)}{{\cal B}\left(B_{s}^{0} \to J/\psi h_{1}(1170)\right)} \;, \qquad
R_{sd}^{\rm SO}[h_{1}(1415)] \equiv \frac{{\cal B}\left(B_{s}^{0} \to J/\psi h_{1}(1415)\right)}{{\cal B}\left(B_{d}^{0} \to J/\psi h_{1}(1415)\right)} \;,
\eeq
can also be used to constrain the absolute value of mixing angle $\theta$. Our numerical results are
\beq
R_{ds}^{\rm SO}[h_{1}(1170)]\approx 6.38_{-0.89}^{+1.17}\;, \qquad R_{sd}^{\rm SO}[h_{1}(1415)]\approx 13.18_{-1.92}^{+2.02} \times 10^{2}\;,
\eeq
in which, all errors arising from various input parameters have been added in quadrature.

Now, we turn to investigate the polarization fractions of the $B^{0} \to J/\psi h_{1}$ decays.
Based on the helicity amplitudes given in Eqs.~\eqref{eq:DecAmp-psih11-d}-\eqref{eq:DecAmp-psih14-s},
we can equivalently define a set of transversity amplitudes as follows,
\beq
{\cal A}_{L}&=&m_{B^{0}}^{2}{\cal A}_{L}\;,
\qquad
{\cal A}_{\parallel}=\sqrt{2}m_{B^{0}}^{2}{\cal A}_{N}\;,
\qquad
{\cal A}_{\perp}=m_{J/\psi}m_{h_{1}}\sqrt{2({\kappa}^{2}-1)}{\cal A}_{T}\;,
\eeq
for the longitudinal, parallel and perpendicular polarization states, respectively, with the ratio $\kappa=P_{2} \cdot P_{3} /(m_{J/\psi} m_{h_{1}})$.
Then, we can define the polarization fractions as
\beq
f_{L,\parallel,\perp}\equiv \frac{|{\cal A}_{L,\parallel,\perp}|^{2}}{|{\cal A}_{L}|^{2}+|{\cal A}_{\parallel}|^{2}+|{\cal A}_{\perp}|^{2}}\;, \qquad
f_{T}\equiv \frac{|{\cal A}_{\parallel}|^{2}+|{\cal A}_{\perp}|^{2}}{|{\cal A}_{L}|^{2}+|{\cal A}_{\parallel}|^{2}+|{\cal A}_{\perp}|^{2}}=f_{\parallel}+f_{\perp}\;,
\label{eq:fl3}
\eeq
which are obviously satisfy the relation $f_{L}+f_{\parallel}+f_{\perp}=f_{L}+f_{T}=1$.
In addition, the relative phases $\phi_{\parallel}$ and $\phi_{\perp}$ (in units of rad) are
defined as
\beq
\phi_{\parallel}&=& {\rm arg}\frac{{\cal A}_{\parallel}}{{\cal A}_{L}}\;,
\qquad
\phi_{\perp}= {\rm arg}\frac{{\cal A}_{\perp}}{{\cal A}_{L}}\;.
\eeq

\begin{table}[h]
\renewcommand{\arraystretch}{1.10}	
\caption{ Theoretical predictions for the polarization observables of the $B_{d}^{0} \to J/\psi h_{1}$ decays.}
\label{tab:three pola-d}
\begin{center}
\begin{tabular}[t]{c|c|c}
\hline  \hline
\makebox[0.15\textwidth][c]{Observables}
&\makebox[0.30\textwidth][c]{$B_{d}^{0} \to J/\psi h_{1}(1170)$}
&\makebox[0.30\textwidth][c]{$B_{d}^{0} \to J/\psi h_{1}(1415)$} \\
\hline \hline
{$f_{L}(\%)$}
&$\begin{array}{ll} 81.3_{-0.3-0.1-1.8-4.5-0.2}^{+0.3+0.1+1.3+3.6+0.2} \end{array} $
&$\begin{array}{ll} 80.3_{-0.6-0.1-1.7-3.8-0.4}^{+0.6+0.1+0.9+3.1+0.3} \end{array} $
\\
$f_{\parallel}(\%)$
&$\begin{array}{ll} 11.8_{-0.2-0.1-0.6-2.1-0.1}^{+0.2+0.1+0.9+2.7+0.1} \end{array} $
&$\begin{array}{ll} 13.5_{-0.5-0.1-0.3-2.0-0.1}^{+0.5+0.1+1.1+2.4+0.2} \end{array} $
\\
$f_{\perp}(\%)$
&$\begin{array}{ll} \;\;6.9_{-0.1-0.0-0.7-1.5-0.1}^{+0.1+0.0+0.9+1.9+0.1} \end{array} $
&$\begin{array}{ll} \;\;6.2_{-0.2-0.0-0.6-1.1-0.1}^{+0.2+0.0+0.7+1.5+0.2} \end{array} $	
\\ \hline
$\phi_{\parallel}({\rm rad})$
&$\begin{array}{ll}  3.16_{-0.02-0.00-0.13-0.10-0.00}^{+0.02+0.00+0.11+0.12+0.01} \end{array} $
&$\begin{array}{ll}  3.36_{-0.03-0.01-0.10-0.11-0.01}^{+0.02+0.01+0.07+0.13+0.01} \end{array} $
\\
$\phi_{\perp}({\rm rad})$
&$\begin{array}{ll}  3.10_{-0.01-0.00-0.12-0.10-0.00}^{+0.01+0.00+0.09+0.13+0.00} \end{array} $
&$\begin{array}{ll}  3.30_{-0.01-0.01-0.09-0.11-0.01}^{+0.01+0.01+0.06+0.14+0.01} \end{array} $
\\ \hline
${A}_{\rm CP}^{\rm dir,L}(10^{-2})$
&$\begin{array}{ll}  -0.81_{-0.05-0.02-0.02-0.09-0.44}^{+0.04+0.02+0.02+0.06+0.40} \end{array} $
&$\begin{array}{ll}  -1.05_{-0.06-0.03-0.05-0.10-0.44}^{+0.06+0.03+0.07+0.08+0.43} \end{array} $
\\
${A}_{\rm CP}^{\rm dir,\parallel}(10^{-2})$
&$\begin{array}{ll} -1.17_{-0.07-0.03-0.07-0.10-0.45}^{+0.07+0.03+0.07+0.12+0.42} \end{array} $
&$\begin{array}{ll} -1.27_{-0.09-0.04-0.09-0.12-0.49}^{+0.08+0.04+0.10+0.14+0.45} \end{array} $
\\
${A}_{\rm CP}^{\rm dir,\perp}(10^{-2})$
&$\begin{array}{ll} -0.99_{-0.07-0.02-0.09-0.09-0.42}^{+0.06+0.02+0.07+0.11+0.39} \end{array} $
&$\begin{array}{ll} -1.18_{-0.09-0.03-0.14-0.10-0.44}^{+0.09+0.03+0.13+0.15+0.43} \end{array} $
\\ \hline \hline
\end{tabular}
\end{center}
\end{table}
\begin{table}[h]
\renewcommand{\arraystretch}{1.15}	
\caption{ Same as Table~\ref{tab:three pola-d} but for $B_{s}^{0} \to J/\psi h_{1}$ decays.}
\label{tab:three pola-s}
\begin{center}	
\begin{tabular}[t]{c|c|c}
\hline  \hline
\makebox[0.15\textwidth][c]{Decay modes}
&\makebox[0.30\textwidth][c]{$B_{s}^{0} \to J/\psi h_{1}(1170)$}
&\makebox[0.30\textwidth][c]{$B_{s}^{0} \to J/\psi h_{1}(1415)$} \\
\hline \hline
{$f_{L}(\%)$}
&$\begin{array}{ll} 87.6_{-0.2-0.1-2.7-5.7-0.2}^{+0.3+0.1+2.7+4.5+0.2} \end{array} $
&$\begin{array}{ll} 89.4_{-0.3-0.0-2.8-5.9-0.1}^{+0.3+0.0+2.6+4.3+0.1} \end{array} $
\\
$f_{\parallel}(\%)$
&$\begin{array}{ll} \;\;7.1_{-0.2-0.1-1.5-2.6-0.1}^{+0.2+0.1+1.6+3.3+0.1} \end{array} $
&$\begin{array}{ll} \;\;6.7_{-0.2-0.0-1.6-2.7-0.1}^{+0.2+0.0+1.8+3.7+0.1} \end{array} $
\\
$f_{\perp}(\%)$
&$\begin{array}{ll} \;\;5.3_{-0.1-0.0-1.2-1.9-0.1}^{+0.1+0.1+1.1+2.4+0.1} \end{array} $
&$\begin{array}{ll} \;\;3.9_{-0.1-0.0-1.0-1.6-0.1}^{+0.1+0.0+1.0+2.2+0.1} \end{array} $
\\ \hline
$\phi_{\parallel}({\rm rad})$
&$\begin{array}{ll}  2.61_{-0.01-0.01-0.19-0.07-0.01}^{+0.01+0.01+0.14+0.11+0.01} \end{array} $
&$\begin{array}{ll}  2.74_{-0.01-0.00-0.19-0.11-0.00}^{+0.01+0.00+0.16+0.16+0.01} \end{array} $
\\
$\phi_{\perp}({\rm rad})$
&$\begin{array}{ll}  2.61_{-0.01-0.01-0.18-0.08-0.01}^{+0.01+0.01+0.15+0.11+0.01} \end{array} $
&$\begin{array}{ll}  2.74_{-0.01-0.00-0.19-0.11-0.00}^{+0.01+0.00+0.16+0.17+0.01} \end{array} $
\\ \hline
${A}_{\rm CP}^{\rm dir,L}(10^{-4})$
&$\begin{array}{ll}  1.65_{-0.10-0.02-0.27-0.01-1.46}^{+0.02+0.03+0.48+0.01+1.86} \end{array} $
&$\begin{array}{ll}  1.37_{-0.01-0.01-0.09-0.00-1.66}^{+0.00+0.01+0.10+0.00+1.97} \end{array} $
\\
${A}_{\rm CP}^{\rm dir,\parallel}(10^{-4})$
&$\begin{array}{ll} 1.64_{-0.14-0.06-0.20-0.71-1.18}^{+0.15+0.08+0.34+0.27+1.12} \end{array} $
&$\begin{array}{ll} 0.52_{-0.07-0.00-0.00-0.96-1.20}^{+0.09+0.00+0.11+0.63+1.39} \end{array} $
\\
${A}_{\rm CP}^{\rm dir,\perp}(10^{-4})$
&$\begin{array}{ll} 1.55_{-0.09-0.06-0.11-0.72-0.97}^{+0.09+0.08+0.21+0.29+1.11} \end{array} $
&$\begin{array}{ll} 0.58_{-0.09-0.01-0.05-0.97-1.21}^{+0.09+0.01+0.07+0.59+1.37} \end{array} $
\\ \hline \hline
\end{tabular}
\end{center}
\end{table}

\begin{figure}[h]
\begin{center}
\includegraphics[scale=0.50]{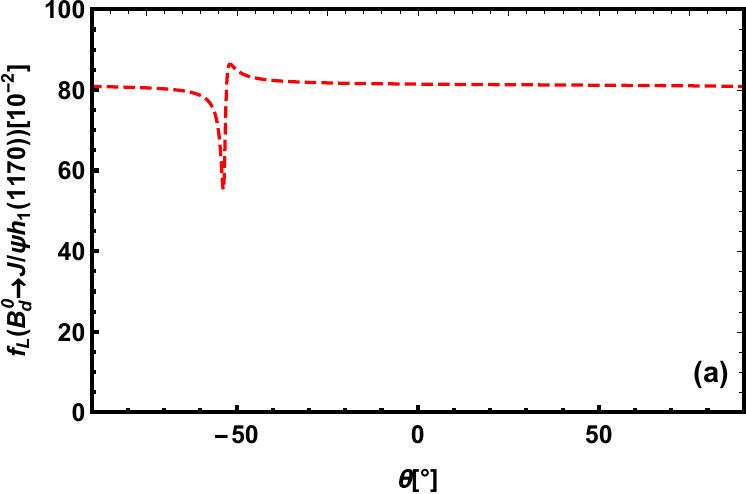}\quad
\includegraphics[scale=0.50]{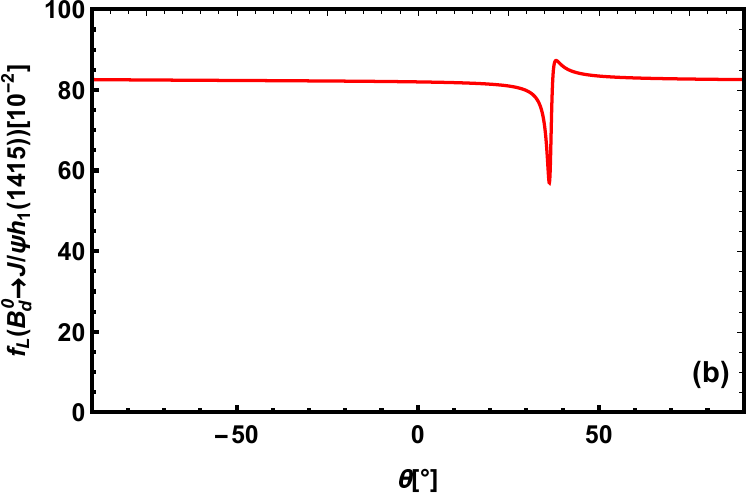}\quad
\includegraphics[scale=0.50]{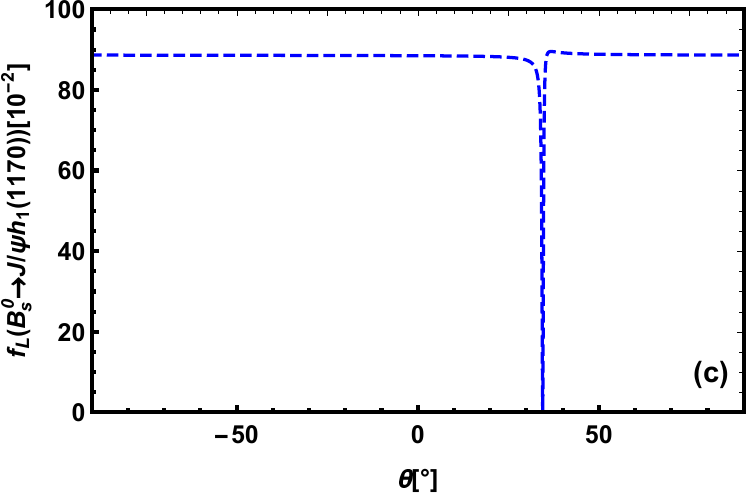}\quad
\includegraphics[scale=0.50]{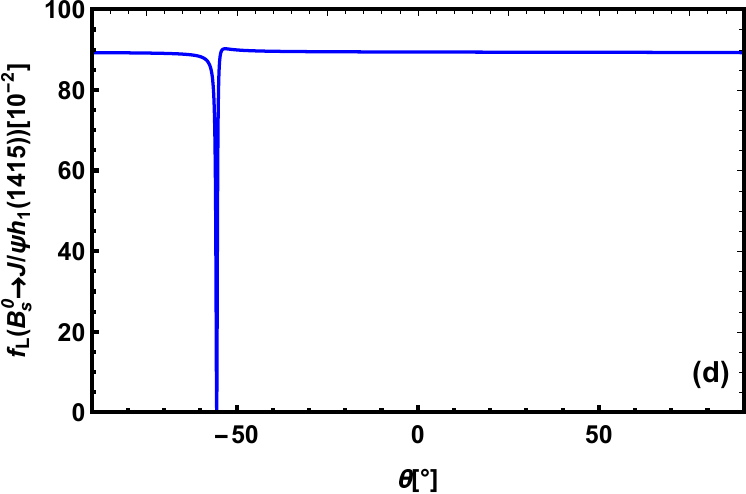}
\caption{(Color online) The dependence of ${f_L}(B^{0} \to J/\psi h_{1})$ on the mixing angle $\theta$ in the PQCD approach.
 }
\label{fig:figfL}
\end{center}\vspace{-0.5cm}
\end{figure}

Our numerical results for the polarization observables of $B^{0} \to J/\psi h_{1}$ decays
are given in Tables~\ref{tab:three pola-d} and \ref{tab:three pola-s}.
Adding the errors in quadrature, the results of polarization fractions can be simplified as
\beq
f_{L}\left(B_{d}^{0} \to J/\psi h_{1}(1170)\right)&=&(81.3_{-4.9}^{+3.9})\%\;, \qquad
f_{T}\left(B_{d}^{0} \to J/\psi h_{1}(1170)\right)=(18.7_{-2.7}^{+3.5})\%\;, \\
f_{L}\left(B_{d}^{0} \to J/\psi h_{1}(1415)\right)&=&(80.3_{-4.3}^{+3.2})\%\;,  \qquad
f_{T}\left(B_{d}^{0} \to J/\psi h_{1}(1415)\right)=(19.7_{-2.4}^{+3.1})\%\;,
\eeq
and
\beq
f_{L}\left(B_{s}^{0} \to J/\psi h_{1}(1170)\right)&=&(87.6_{-6.3}^{+5.3})\%\;,  \qquad
f_{T}\left(B_{s}^{0} \to J/\psi h_{1}(1170)\right)=(12.4_{-3.7}^{+4.5})\%\;,  \\
f_{L}\left(B_{s}^{0} \to J/\psi h_{1}(1415)\right)&=&(89.4_{-6.5}^{+5.0})\%\;,  \qquad
f_{T}\left(B_{s}^{0} \to J/\psi h_{1}(1415)\right)=(10.6_{-3.6}^{+4.7})\%\;,
\eeq
It can be found that $f_L(B^{0} \to J/\psi h_{1})$ are generally larger than $80\%$,
which indicate that $B^{0} \to J/\psi h_{1}$ decays are dominated by the longitudinal
contributions. The dependence of ${f_L}(B^{0} \to J/\psi h_{1})$ on the mixing angle $\theta$
is shown in Fig.~\ref{fig:figfL}. It can be found that the polarization fractions generally
are not sensitive to the value of $\theta$, except at $\theta\sim 35^\circ$ for
$B_{d}^{0} \to J/\psi h_{1}(1415)$ and $B_{s}^{0} \to J/\psi h_{1}(1170)$ decays
and at $\theta\sim -55^\circ$ for $B_{d}^{0} \to J/\psi h_{1}(1170)$ and $B_{s}^{0} \to J/\psi h_{1}(1415)$
decays. Thus, the polarization would present a very strong constraint on $\theta$ if a relatively
small $f_L(B^{0} \to J/\psi h_{1})$ is observed by the experiments.
\begin{figure}[h]
\begin{center}\vspace{0.8cm}
\includegraphics[scale=0.50]{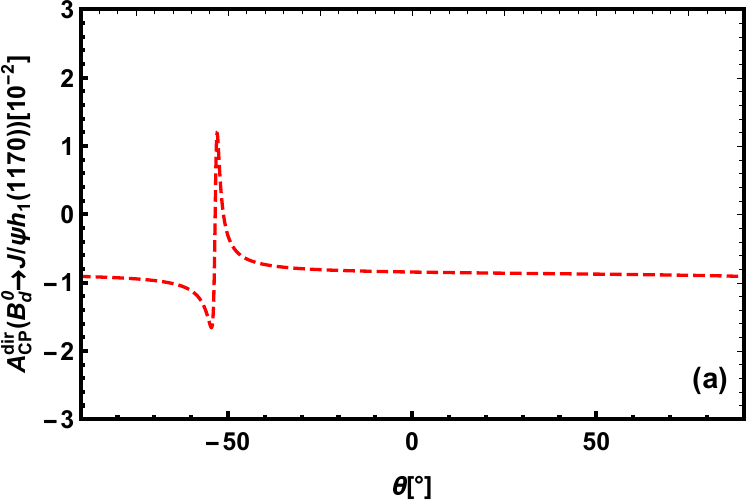}\quad
\includegraphics[scale=0.50]{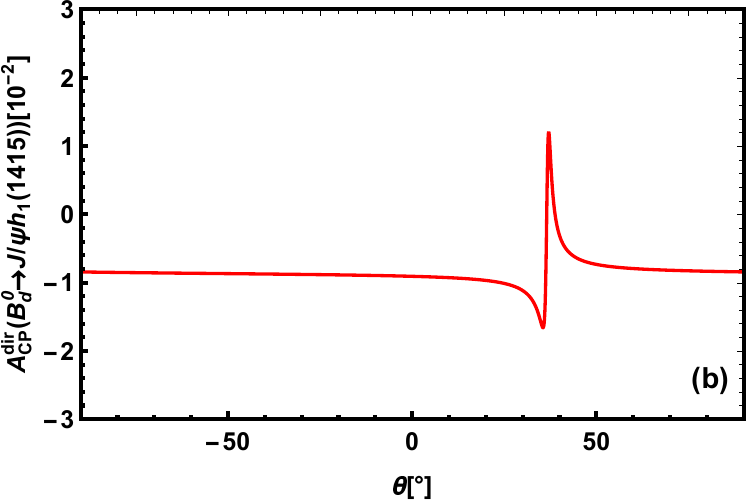}\quad
\includegraphics[scale=0.50]{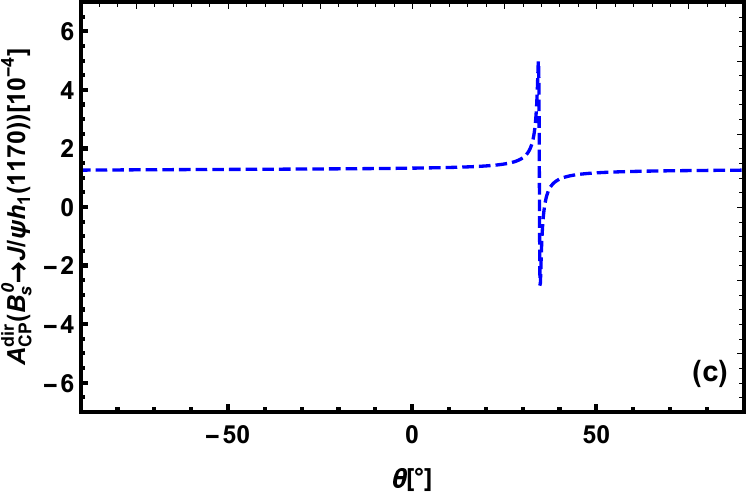}\quad
\includegraphics[scale=0.50]{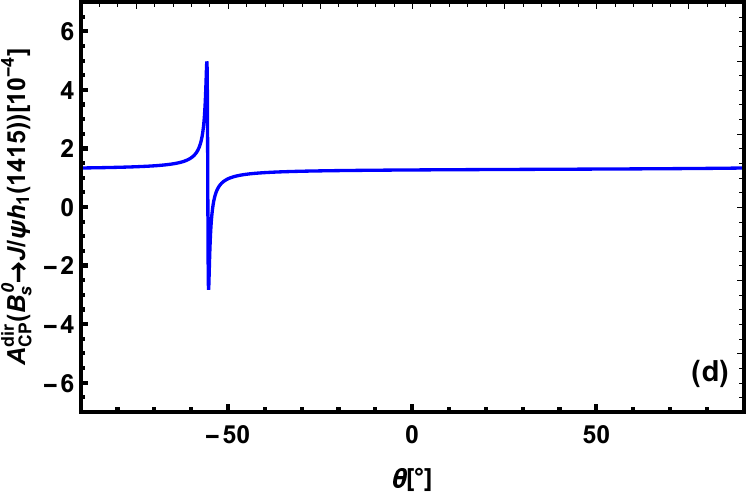}
\caption{(Color online) The dependence of ${A_{\rm CP}^{\rm dir}}(B^{0} \to J/\psi h_{1})$ on the mixing angle $\theta$ in the PQCD approach. }
\label{fig:figACP}
\end{center}\vspace{-0.5cm}
\end{figure}

The direct {\it CP} asymmetry $A_{\rm CP}^{\rm dir}$ of $B^0 \to J/\psi h_1$ decays is defined as
\beq
A_{\rm CP}^{\rm dir}
\equiv  \frac{|\overline{\cal A}(\overline{B}^0\to \overline f)|^2 - |{\cal A}(B^0\to f)|^2}
{|\overline{\cal A}(\overline{B}^0\to \overline f)|^2+|{\cal A}(B^0\to f)|^2}\;,
\label{eq:acp1}
\eeq
where ${\cal A}$ represents the decay amplitudes of $B^0 \to J/\psi h_{1}$, while $\overline{\cal A}$ describe the corresponding charge conjugation ones. Meanwhile, according to Ref.~\cite{Beneke:2006hg}, the direct {\it CP} asymmetries in each polarization can also be studied as
\beq
A_{\rm CP}^{\rm dir,\alpha} = \frac{\overline{f}_{\alpha}-f_{\alpha}}{\overline{f}_{\alpha}+f_{\alpha}} \quad (\alpha=L,\parallel,\perp)\;,
\eeq
where $\overline{f}_{\alpha}$ is the polarization fraction for the corresponding $\overline{B}^{0}$ decays in Eq.\eqref{eq:fl3}. 
Using Eq.\eqref{eq:acp1}, our numerical predictions for the direct {\it CP} asymmetries of the $B^{0} \to J/\psi h_{1}$ decays in the PQCD approach are
\beq
\label{eq:cp-dir-d11}
\resizebox{0.46\textwidth}{!}{$
A_{\rm CP}^{\rm dir}\left(B_{d}^{0} \to J/\psi h_{1}(1170)\right)=-0.86_{-0.44}^{+0.41} \times 10^{-2},$} \quad
\resizebox{0.46\textwidth}{!}{$
A_{\rm CP}^{\rm dir}\left(B_{d}^{0} \to J/\psi h_{1}(1415)\right)=-1.09_{-0.46}^{+0.45}  \times 10^{-2},$} \quad \\
\resizebox{0.46\textwidth}{!}{$
A_{\rm CP}^{\rm dir}\left(B_{s}^{0} \to J/\psi h_{1}(1170)\right)=1.64_{-1.44}^{+1.81} \times 10^{-4}, $}\quad
\resizebox{0.46\textwidth}{!}{$
A_{\rm CP}^{\rm dir}\left(B_{s}^{0} \to J/\psi h_{1}(1415)\right)=1.28_{-1.62}^{+1.91} \times 10^{-4}. $}\quad
\label{eq:cp-mix-s14}
\eeq
All the errors arising from various parameters in the above-mentioned have been added in quadrature.
The dependence of ${A_{\rm CP}^{\rm dir}}(B^{0} \to J/\psi h_{1})$ on the mixing angle $\theta$ is
shown in Fig.~\ref{fig:figACP}. It can be found that the direct {\it CP} asymmetries are too small to
be observed, although they are relatively sensitive to $\theta$ at some specific range. 

All of the aforementioned theoretical predictions in the PQCD approach are expected to be tested
by LHCb, Belle-II and proposed CEPC experiments in the future. The relevant predictions for these
experimental observables would be helpful to explore the dynamics involved in the $B^{0} \to J/\psi h_{1}$
decays and to identify the inner structure or the components of the $h_{1}$ states.

\section{Conclusions and summary}
\label{sec04}
In this paper, we have calculated the $B^{0} \to J/\psi h_{1}$ decays for the first time by using PQCD approach,
where $h_{1}=h_{1}(1170)$ and $h_{1}(1415)$ are treated as the mixtures of $\tilde{h}_{1}$ and $\tilde{h}_{8}$
with mixing angle $\theta$ in the singlet-octet basis. The observables of these decays are predicted. The NLO order
corrections are considered in the calculation because the vertex corrections with the NLO Wilson coefficients
contribute significantly to the color-suppressed decay modes. We take $\theta=29.5^{\circ}$
as default input, and the effects of $\theta$ on the observables are discussed in detail.
Our conclusions can be summarized briefly as following,
\begin{itemize}
\item
The $B^{0} \to J/\psi h_{1}$ decays have relatively large branching fractions,
which are generally at the order of ${\cal O}(10^{-6}\sim10^{-3})$, and thus are possible
to be measured by the LHCb and Belle-II experiments in the near future. The measurements
with high precision can provide useful information for basic nature of $h_{1}(1170)$
and $h_{1}(1415)$.	

\item
Further considering the secondary decays, $(h_{1}(1415),f_{1}(1420)) \to K_{S}^{0}K^{+}\pi^{-}$,
and comparing them with the LHCb data for ${\cal B}(B_{s}^{0} \to J/\psi K_{S}^{0}K^{+}\pi^{-})$,
we find that the $h_{1}(1415)$ resonance serves possibly as a contributing state in the $m(K_{S}^{0}K^{\pm}\pi^{\mp})$ distribution.

\item
The branching fractions of $B^{0} \to J/\psi h_{1}$ decays are very sensitive
to the mixing angle, and thus can be used to test the values of $\theta$.
Several interesting ratios between the branching fractions of $B^{0} \to J/\psi h_{1}$
decays, such as $R_{d}^{\rm SO}$, $R_{s}^{\rm SO}$, $R_{ds}^{\rm SO}$, can effectively avoid
large theoretical errors caused by the hadronic inputs, and thus would provide much
stronger constraints on $\theta$.

\item
All of the $B^{0} \to J/\psi h_{1}$ decays are generally dominated by the longitudinal
polarization contributions, namely, $f_{L}(B^{0} \to J/\psi h_{1})>80\%$, with default
input $\theta=29.5^{\circ}$. The longitudinal fractions can be significantly reduced when
some specific values, $\theta\sim  35^\circ\,,-55^\circ$, are taken. Therefore, the polarization
fractions of $B^{0} \to J/\psi h_{1}$ decays can provide restrict constraint on $\theta$.	

\item
Maybe the direct {\it CP} asymmetries of $B^{0} \to J/\psi h_{1}$ decays are too small to be observed
soon even if the effect of $\theta$ is considered.
\end{itemize}

%
%
\begin{acknowledgments}
The work is supported by the National Natural Science Foundation of China (Grant Nos. 12275067, 11875033),
Science and Technology R$\&$D Program Joint Fund Project of Henan Province  (Grant No.225200810030),
Excellent Youth Foundation of Henan Province (Grant no. 212300410010),
 and National Key R$\&$D Program of China (Grant No. 2023YFA1606000), and Science and Technology Innovayion Leading Talent Support Program of Henan Province.
\end{acknowledgments}
%
%
\begin{appendix}
\section{Specific derivation of angle calculations}
\label{sec:app1}
Under the basis $\tilde{h}_{1}$ and $\tilde{h}_{8}$, we can write the mass matrix as follows~\cite{Cheng:2011pb,Du:2021zdg},
\beq
\left( \begin{array}{cc}
	\langle \tilde{h}_{1}|H^{2}|\tilde{h}_{1}\rangle & \langle \tilde{h}_{1}|H^{2}|\tilde{h}_{8}\rangle \\
	\langle \tilde{h}_{8}|H^{2}|\tilde{h}_{1}\rangle & \langle \tilde{h}_{8}|H^{2}|\tilde{h}_{8}\rangle \end{array}\right)=
\left( \begin{array}{cc}
	m_{\tilde{h}_{1}}^{2} & m_{\tilde{h}_{18}}^{2} \\
	m_{\tilde{h}_{18}}^{2} & m_{\tilde{h}_{8}}^{2} \end{array}\right)\;.
\eeq
The mixing angle $\theta$ can be derived by diagonalizing the mass matrix, and the mass matrix diagonalized according to the following relation,
\beq
R(\theta)M^{2}R(\theta)^{-1}=M_{diag.}^{2}\;.
\eeq
Therefore, one can obtains the physical masses of $h_{1}(1170)$ and $h_{1}(1415)$ states,
\beq
\left( \begin{array}{cc}
	m_{h_{1}(1170)}^{2} & 0 \\
	0 & m_{h_{1}(1415)}^{2} \end{array}\right)=
\left( \begin{array}{cc}
	\,\,\cos{\theta} & \,\,\sin{\theta} \\
	-\sin{\theta} & \,\,\cos{\theta} \end{array} \right)
\left( \begin{array}{cc}
	m_{\tilde{h}_{1}}^{2} & m_{\tilde{h}_{18}}^{2} \\
	m_{\tilde{h}_{18}}^{2} & m_{\tilde{h}_{8}}^{2} \end{array}\right)
\left( \begin{array}{cc}
	\,\,\cos{\theta} & -\sin{\theta} \\
	\,\,\sin{\theta} & \,\,\cos{\theta} \end{array} \right)\;.
\eeq
Then, we have
\beq
\label{eq:A5}
m_{h_{1}(1170)}^{2} \cos^{2}\theta &=& m_{\tilde{h}_{1}}^{2} \cos^{2}\theta+m_{\tilde{h}_{18}}^{2}\sin\theta\cos\theta\;, \\
m_{h_{1}(1415)}^{2} \sin^{2}\theta &=& m_{\tilde{h}_{1}}^{2} \sin^{2}\theta-m_{\tilde{h}_{18}}^{2}\cos\theta \sin\theta\; ,
\eeq
\beq
m_{h_{1}(1170)}^{2} \sin^{2}\theta &=& m_{\tilde{h}_{18}}^{2} \cos\theta\sin\theta+m_{\tilde{h}_{8}}^{2}\sin^{2}\theta \;, \\
m_{h_{1}(1415)}^{2} \cos^{2}\theta &=& -m_{\tilde{h}_{18}}^{2} \sin\theta\cos \theta+m_{\tilde{h}_{8}}^{2}\cos^{2}\theta \; .
\label{eq:A8}
\eeq
By utilizing Eqs.~\eqref{eq:A5}-\eqref{eq:A8}, the following relationship can be derived,
\beq
m_{\tilde{h}_{1}}^{2}&=&m_{h_{1}(1415)}^{2} \sin^{2}\theta+m_{h_{1}(1170)}^{2} \cos^{2}\theta\;, \\
m_{\tilde{h}_{8}}^{2}&=&m_{h_{1}(1415)}^{2} \cos^{2}\theta+m_{h_{1}(1170)}^{2} \sin^{2}\theta\;, \\
m_{\tilde{h}_{18}}^{2}&=&-\frac{1}{2}(m_{h_{1}(1415)}^{2}-m_{h_{1}(1170)}^{2})\sin 2\theta\;.
\eeq
After further collation and simplification, the following results can be obtained,
\beq
\cos^{2}\theta&=&\frac{m_{\tilde{h}_{8}}^{2}-m_{h_{1}(1170)}^{2}}{m_{h_{1}(1415)}^{2}-m_{h_{1}(1170)}^{2}}
=\frac{4m_{K_{1B}}^{2}-m_{b_{1}}^{2}-3m_{h_{1}(1170)}^{2}}{3\left(m_{h_{1}(1415)}^{2}-m_{h_{1}(1170)}^{2}\right)}\;, \\
\tan\theta&=&\frac{4m_{K_{1B}}^{2}-m_{b_{1}}^{2}-3m_{h_{1}(1415)}^{2}}{2\sqrt{2}\left (m_{b_{1}}^{2}-K_{1B}^{2}\right )} \;,
\eeq
where, the Gell-Mann$-$Okubo mass relation, $m_{b_{1}}^{2}+3m_{\tilde{h}_{8}}^{2}=4m_{K_{1B}}^{2}$.

\section{Meson wave functions and distribution amplitudes}
\label{sec:app2}
Notice that, the process of $B$-meson decays into $J/\psi$ plus light hadrons have been studied in the PQCD approach at the NLO accuracy~\cite{Chen:2005ht,Yao:2022zom,Liu:2013nea}. Hence, in this article, the same wave functions and the related distribution amplitudes for the heavy $B^{0}$ and $J/\psi$ mesons will not be presented one by one here and could be found in Ref.~\cite{Yao:2022zom}.

For the light axial-vector meson $h_{1}$, the wave functions associated with the light-cone distribution amplitudes at both longitudinal and transverse polarizations have been given in the QCD sum rule up to twist-3.
Therefore, the wave function could be written as~\cite{Yang:2007zt,Li:2009tx},
\beq
\Phi^{L}_{A}(x) &=&  \frac{1}{\sqrt{2 N_c}}\gamma_5
\biggl\{ m_{A}\, {\epsl}_L \,\phi_{A}(x)  +
{\epsl}_L \, \psl\,\phi_{A}^t(x)  + m_{A}\, \phi_{A}^s(x) \biggr\}_{\alpha\beta}\;,
\label{eq:wf-h1-L}
\\
\Phi^{T}_{A}(x) &=&  \frac{ 1}{\sqrt{2 N_c}} \gamma_5
\biggl\{ m_{A}\, {\epsl}_T\, \phi_{A}^v(x) +
{\epsl}_T\, \psl\, \phi_{A}^T(x)+m_{A}
i\epsilon_{\mu\nu\rho\sigma}\gamma_5\gamma^\mu \epsilon_T^{\nu}
n^\rho v^\sigma \phi_{A}^a(x) \biggr\}_{\alpha\beta}\;,
\label{eq:wf-h1-T}
\eeq
where $\epsilon_{L}$ and $\epsilon_{T}$ are the longitudinal and transverse polarization vectors of $h_{1}$ meson,
$x$ denotes the momentum fraction carried by quark in $h_{1}$,
$n=(1,0,{\bf 0}_T)$ and $v=(0,1,{\bf 0}_T)$ are dimensionless lightlike vectors,
the Levi-Civit$\grave{a}$ tensor $\epsilon^{\mu\nu\alpha\beta}$
is conventionally taken as $\epsilon^{0123}=1$. With the twist-2 light-cone distribution amplitudes, i.e., $\phi_{A}(x)$ and $\phi_{A}^T(x)$, can be expanded as the Gegenbauer polynomials~\cite{Li:2009tx},
\beq
\phi_{A}(x)&=&\frac{3f_{A}}{\sqrt{2N_c}} x(1-x) \biggl[3 a_{1}^\parallel \left(2x-1\right) \biggr]\;,
\label{eq:da-fh1-L}
\\
\phi_{A}^T(x)&=& \frac{3f_{A}}{\sqrt{2N_c}} x(1-x) \biggl[1+a_{2}^\perp\; \frac{3}{2}\left(5\left(2x-1\right)^{2}-1\right)  \biggr] \;.
\label{eq:da-fh1-T}
\eeq
And the twist-3 light-cone distribution amplitudes will be used in the following form~\cite{Li:2009tx},
\beq
\phi_{A}^s(x)&=&\frac{3f_{A}}{2\sqrt{2N_c}} \frac{d}{dx}\biggl[ x(1-x) \biggr]\;,
\label{eq:da-fh1-s} \qquad
\phi_{A}^t(x)=\frac{3f_{A}}{2\sqrt{2N_c}} \biggl[ (2x-1)^{2}\biggr]\;,
\label{eq:da-fh1-v}
\\
\phi_{A}^v(x)&=& \frac{3f_{A}}{4\sqrt{2N_c}} \biggl[ a_{1}^{\parallel}(2x-1)^3 \biggr]\;,
\label{eq:da-fh1-t} \qquad
\phi_{A}^a(x) = \frac{3f_{A}}{4\sqrt{2N_c}} \frac{d}{dx}\biggl[ x(1-x)\left(a_{1}^{\parallel}(2x-1)\right) \biggr]\;,
\label{eq:da-fh1-a}
\eeq
where $f_{A}$ is the decay constant of the single-octet states $f_{\tilde{h}_{1,8}}$ and the Gegenbauer moments $a_{1}^\parallel$ and $a_{2}^\perp$ in Eqs.~\eqref{eq:da-fh1-L}-\eqref{eq:da-fh1-a}at the renormalization scale $\mu$=1 ${\rm GeV}$ are as follows~\cite{Yang:2007zt},
\beq
a_1^\parallel &=& \left\{ \begin{array}{ll}
	-1.95^{+0.35}_{-0.35} \;\;\; \left(\;{\rm for}\; \tilde{h}_{8}\;\right),&  \\
	-2.00^{+0.35}_{-0.35} \;\;\; \left(\;{\rm for}\; \tilde{h}_{1}\;\right),&   \\ \end{array} \right.
\qquad
a_2^\perp = \left\{ \begin{array}{ll}
	0.14^{+0.22}_{-0.22} \;\;\; \left(\;{\rm for}\; \tilde{h}_{8}\;\right),&  \\
	0.18^{+0.22}_{-0.22} \;\;\; \left(\;{\rm for}\; \tilde{h}_{1}\;\right).&   \\ \end{array} \right.
\label{eq:Geng-h1}
\eeq

\end{appendix}
\newpage


\begin{thebibliography}{99}
\bibitem{Belle:2001zzw}
K.~Abe \textit{et al.} [Belle],
Phys. Rev. Lett. \textbf{87}, 091802 (2001).

	
\bibitem{Belle:2001rjp}
K.~Abe \textit{et al.} [Belle],
Phys. Rev. Lett. \textbf{87}, 161601 (2001).


\bibitem{Burakovsky:1997dd}
L.~Burakovsky and J.~T.~Goldman,
Phys. Rev. D \textbf{56}, R1368-R1372 (1997).


\bibitem{Cheng:2011pb}
H.~Y.~Cheng,
Phys. Lett. B \textbf{707}, 116-120 (2012).


\bibitem{Chen:2015iqa}
K.~Chen, C.~Q.~Pang, X.~Liu and T.~Matsuki,
Phys. Rev. D \textbf{91}, 074025 (2015).


\bibitem{ParticleDataGroup:2022pth}
R.~L.~Workman \textit{et al.} [Particle Data Group],
PTEP \textbf{2022}, 083C01 (2022).


\bibitem{HFLAV:2022esi}
Y.~S.~Amhis \textit{et al.} [HFLAV],
Phys. Rev. D \textbf{107}, 052008 (2023).


\bibitem{Liang:2019vhf}
W.~H.~Liang, S.~Sakai and E.~Oset,
Phys. Rev. D \textbf{99}, 094020 (2019).


\bibitem{Liu:2010epa}
X.~Liu and Z.~J.~Xiao,
J. Phys. G \textbf{38}, 035009 (2011).


\bibitem{Liu:2010da}
X.~Liu and Z.~J.~Xiao,
Phys. Rev. D \textbf{81}, 074017 (2010).


\bibitem{Du:2021zdg}
M.~C.~Du and Q.~Zhao,
Phys. Rev. D \textbf{104}, 036008 (2021).


\bibitem{Du:2022nno}
M.~C.~Du, Y.~Cheng and Q.~Zhao,
Phys. Rev. D \textbf{106}, 054019 (2022).


\bibitem{BESIII:2015vfb}
M.~Ablikim \textit{et al.} [BESIII],
Phys. Rev. D \textbf{91}, 112008 (2015).


\bibitem{BESIII:2018ede}
M.~Ablikim \textit{et al.} [BESIII],
Phys. Rev. D \textbf{98}, 072005 (2018).


\bibitem{BESIII:2022zel}
M.~Ablikim \textit{et al.} [BESIII],
Phys. Rev. D \textbf{105}, 072002 (2022).


\bibitem{ParticleDataGroup:2024cfk}
S.~Navas \textit{et al.} [Particle Data Group],
Phys. Rev. D \textbf{110}, 030001 (2024).


\bibitem{Cheng:2013cwa}
H.~Y.~Cheng,
PoS \textbf{Hadron2013}, 090 (2013).


\bibitem{Suzuki:1993yc}
M.~Suzuki,
Phys. Rev. D \textbf{47}, 1252-1255 (1993).


\bibitem{Blundell:1995au}
H.~G.~Blundell, S.~Godfrey and B.~Phelps,
Phys. Rev. D \textbf{53}, 3712-3722 (1996).


\bibitem{Burakovsky:1997ci}
L.~Burakovsky and J.~T.~Goldman,
Phys. Rev. D \textbf{57}, 2879-2888 (1998).


\bibitem{Dag:2012zz}
H.~Dag, A.~Ozpineci, A.~Cagil and G.~Erkol,
J. Phys. Conf. Ser. \textbf{348}, 012012 (2012).


\bibitem{Divotgey:2013jba}
F.~Divotgey, L.~Olbrich and F.~Giacosa,
Eur. Phys. J. A \textbf{49}, 135 (2013).


\bibitem{Gao:2019jme}
Y.~Gao, Y.~Zhang, B.~Zheng, Z.~H.~Zhang, W.~Yan and X.~Li,
[arXiv:1911.06967 [hep-ph]].


\bibitem{Yang:2007zt}
K.~C.~Yang,
Nucl. Phys. B \textbf{776}, 187-257 (2007).


\bibitem{Chen:2005ht}
C.~H.~Chen and H.~N.~Li,
Phys. Rev. D \textbf{71}, 114008 (2005).


\bibitem{Li:2007xf}
J.~W.~Li and D.~S.~Du,
Phys. Rev. D \textbf{78}, 074030 (2008).


\bibitem{Li:2012sw}
J.~W.~Li, D.~S.~Du and C.~D.~Lu,
Eur. Phys. J. C \textbf{72}, 2229 (2012).


\bibitem{Liu:2014doa}
X.~Liu and Z.~J.~Xiao,
Phys. Rev. D \textbf{89}, 097503 (2014).


\bibitem{Liu:2019ymi}
X.~Liu, Z.~T.~Zou, Y.~Li and Z.~J.~Xiao,
Phys. Rev. D \textbf{100}, 013006 (2019).


\bibitem{Yao:2022zom}
D.~H.~Yao, X.~Liu, Z.~T.~Zou, Y.~Li and Z.~J.~Xiao,
Eur. Phys. J. C \textbf{83}, 13 (2023).


\bibitem{Cheng:2000kt}
H.~Y.~Cheng and K.~C.~Yang,
Phys. Rev. D \textbf{63}, 074011 (2001).

\bibitem{Sun:2013dla}
J.~Sun, Z.~Xiong, Y.~Yang and G.~Lu,
Eur. Phys. J. C \textbf{73}, 2437 (2013).


\bibitem{Song:2002gw}
Z.~z.~Song, C.~Meng and K.~T.~Chao,
Eur. Phys. J. C \textbf{36}, 365-370 (2004).


\bibitem{Meng:2005er}
C.~Meng, Y.~J.~Gao and K.~T.~Chao,
Phys. Rev. D \textbf{87}, 074035 (2013).


\bibitem{Li:2006vq}
H.~n.~Li and S.~Mishima,
JHEP \textbf{03}, 009 (2007).


\bibitem{Beneke:2008pi}
M.~Beneke and L.~Vernazza,
Nucl. Phys. B \textbf{811}, 155-181 (2009).


\bibitem{Colangelo:2010wg}
P.~Colangelo, F.~De Fazio and W.~Wang,
Phys. Rev. D \textbf{83}, 094027 (2011).


\bibitem{Liu:2013nea}
X.~Liu, W.~Wang and Y.~Xie,
Phys. Rev. D \textbf{89}, 094010 (2014).


\bibitem{Wang:2015uea}
W.~F.~Wang, H.~n.~Li, W.~Wang and C.~D.~L\"u,
Phys. Rev. D \textbf{91}, 094024 (2015).


\bibitem{Zhang:2017cbi}
Z.~Q.~Zhang, H.~Guo and S.~Y.~Wang,
Eur. Phys. J. C \textbf{78}, 219 (2018).


\bibitem{Rui:2019yxx}
Z.~Rui, Y.~Li and H.~Li,
Eur. Phys. J. C \textbf{79}, 792 (2019).


\bibitem{Li:2020app}
Y.~Q.~Li, M.~K.~Jia and Z.~Rui,
Chin. Phys. C \textbf{44}, 113104 (2020).


\bibitem{Liu:2018kuo}
X.~Liu, H.~n.~Li and Z.~J.~Xiao,
Phys. Rev. D \textbf{97}, 113001 (2018).


\bibitem{Liu:2020upy}
X.~Liu, H.~n.~Li and Z.~J.~Xiao,
Phys. Lett. B \textbf{811}, 135892 (2020).


\bibitem{Liu:2023kxr}
X.~Liu,
Phys. Rev. D \textbf{108}, 096006 (2023).


\bibitem{Keum:2000ph}
Y.~Y.~Keum, H.~n.~Li and A.~I.~Sanda,
Phys. Lett. B \textbf{504}, 6-14 (2001).


\bibitem{Keum:2000wi}
Y.~Y.~Keum, H.~N.~Li and A.~I.~Sanda,
Phys. Rev. D \textbf{63}, 054008 (2001).


\bibitem{Lu:2000em}
C.~D.~Lu, K.~Ukai and M.~Z.~Yang,
Phys. Rev. D \textbf{63}, 074009 (2001).


\bibitem{Lu:2000hj}
C.~D.~Lu and M.~Z.~Yang,
Eur. Phys. J. C \textbf{23}, 275-287 (2002).

\bibitem{Ali:2007ff}
A.~Ali, G.~Kramer, Y.~Li, C.~D.~Lu, Y.~L.~Shen, W.~Wang and Y.~M.~Wang,
Phys. Rev. D \textbf{76}, 074018 (2007).

\bibitem{Chai:2022ptk}
J.~Chai, S.~Cheng, Y.~h.~Ju, D.~C.~Yan, C.~D.~L\"u and Z.~J.~Xiao,
Chin. Phys. C \textbf{46}, 123103 (2022).



\bibitem{Xiao:2019mpm}
Z.~J.~Xiao, D.~C.~Yan and X.~Liu,
Nucl. Phys. B \textbf{953}, 114954 (2020).


\bibitem{Ren:2023ebq}
J.~L.~Ren, M.~Q.~Li, X.~Liu, Z.~T.~Zou, Y.~Li and Z.~J.~Xiao,
Eur. Phys. J. C \textbf{84}, 358 (2024).


\bibitem{Buchalla:1995vs}
G.~Buchalla, A.~J.~Buras and M.~E.~Lautenbacher,
Rev. Mod. Phys. \textbf{68}, 1125-1144 (1996).


\bibitem{Xiao:2008sw}
Z.~J.~Xiao, Z.~Q.~Zhang, X.~Liu and L.~B.~Guo,
Phys. Rev. D \textbf{78}, 114001 (2008).


\bibitem{Wolfenstein:1983yz}
L.~Wolfenstein,
Phys. Rev. Lett. \textbf{51}, 1945 (1983).


\bibitem{Verma:2011yw}
R.~C.~Verma,
J. Phys. G \textbf{39}, 025005 (2012).


\bibitem{LHCb:2014sli}
R.~Aaij \textit{et al.} [LHCb],
JHEP \textbf{07}, 140 (2014).



\bibitem{Fleischer:2011au}
R.~Fleischer, R.~Knegjens and G.~Ricciardi,
Eur. Phys. J. C \textbf{71}, 1832 (2011).


\bibitem{Beneke:2006hg}
M.~Beneke, J.~Rohrer and D.~Yang,
Nucl. Phys. B \textbf{774}, 64-101 (2007).



\bibitem{Li:2009tx}
R.~H.~Li, C.~D.~Lu and W.~Wang,
Phys. Rev. D \textbf{79}, 034014 (2009).

\end{thebibliography}
\end{document}